\documentclass[article,nojss]{jss}\usepackage[]{graphicx}\usepackage[]{xcolor}
\makeatletter
\def\maxwidth{ %
  \ifdim\Gin@nat@width>\linewidth
    \linewidth
  \else
    \Gin@nat@width
  \fi
}
\makeatother

\usepackage{thumbpdf,lmodern}

\usepackage{amsmath}
\usepackage{amssymb}
\usepackage{makecell}
\usepackage{rotating}

\newcommand{\mat}[1]{\boldsymbol{#1}}         
\newcommand{\vect}[1]{\boldsymbol{#1}}        
\newcommand{\obs}[1]{\mathbf{#1}}             
\newcommand{\argmin}{\mathop{\text{argmin}}}  

\author{Andreas Alfons \\ Erasmus University Rotterdam
   \And N\"{u}fer Y. Ate\c{s} \\ Sabanc{\i} University
   \And Patrick J.F. Groenen \\ Erasmus University Rotterdam}
\Plainauthor{Andreas Alfons, Nufer Y. Ates, Patrick J.F. Groenen}

\title{Robust Mediation Analysis: The \proglang{R} Package \pkg{robmed}}
\Plaintitle{Robust Mediation Analysis: The R Package robmed}
\Shorttitle{Robust Mediation Analysis: The \proglang{R} Package \pkg{robmed}}

\Abstract{
Mediation analysis is one of the most widely used statistical techniques in
the social, behavioral, and medical sciences.  Mediation models allow to study
how an independent variable affects a dependent variable indirectly through one
or more intervening variables, which are called mediators.  The analysis is
often carried out via a series of linear regressions, in which case the
indirect effects can be computed as products of coefficients from those
regressions.  Statistical significance of the indirect effects is typically
assessed via a bootstrap test based on ordinary least-squares estimates.
However, this test is sensitive to outliers or other deviations from normality
assumptions, which poses a serious threat to empirical testing of theory about
mediation mechanisms.  The \proglang{R} package \pkg{robmed} implements a
robust procedure for mediation analysis based on the fast-and-robust bootstrap
methodology for robust regression estimators, which yields reliable results
even when the data deviate from the usual normality assumptions.  Various
other procedures for mediation analysis are included in package \pkg{robmed}
as well.  Moreover, \pkg{robmed} introduces a new formula interface that allows
to specify mediation models with a single formula, and provides various plots
for diagnostics or visual representation of the results.
}

\Plainkeywords{mediation analysis, robust statistics, bootstrap, R}

\Address{
  Andreas Alfons\\
  Econometric Institute\\
  Erasmus School of Economics\\
  Erasmus University Rotterdam\\
  PO Box 1738\\
  3000DR Rotterdam, The Netherlands\\
  E-mail: \email{alfons@ese.eur.nl}\\
  URL: \url{https://personal.eur.nl/alfons/}
}

\IfFileExists{upquote.sty}{\usepackage{upquote}}{}
\begin{document}




\section{Introduction} \label{sec:intro}

\begin{figure}[t]
\begin{center}
\includegraphics[width=0.5\textwidth]{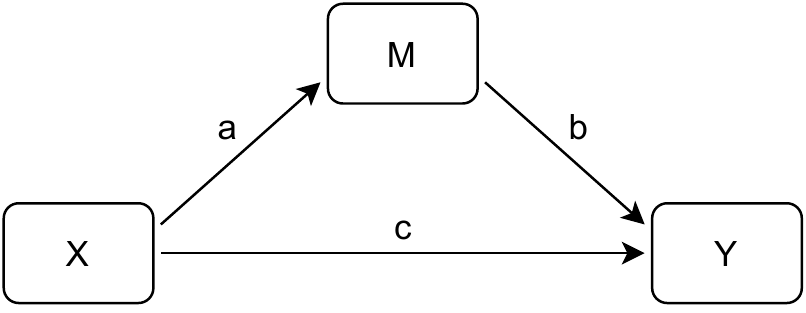}
\end{center}
\caption{Diagram visualizing a simple mediation model.}
\label{fig:simple}
\end{figure}

In the social, behavioral, and medical sciences, mediation analysis is a
popular statistical technique for studying how an independent variable
affects a dependent variable indirectly through an intervening variable
called a mediator.
For instance, \citet{erreygers18} find that poor sleep quality in adolescents
explains cyberbullying through anger, and \citet{gaudiano10} report that the
believability of hallucinations after treatment for psychotic disorders
mediates the relationship between the type of treatment and distress after
treatment.
Figure~\ref{fig:simple} shows a diagram of the mediation model in its
simplest form, which is given by the equations
\begin{align}
M &= i_{1} + a X + e_{1}, \label{eq:simpleM} \\
Y &= i_{2} + b M + c X + e_{2}, \label{eq:simpleY} \\
Y &= i_{3} + c' X + e_{3}, \label{eq:simpleY'}
\end{align}
where $Y$ denotes the dependent variable, $X$ the independent variable, $M$ the
hypothesized mediator, $i_{1}$, $i_{2}$, $i_{3}$, $a$, $b$, $c$, and $c'$ are
regression coefficients to be estimated, and $e_{1}$, $e_{2}$, and $e_{3}$ are
random error terms.  The coefficients $c$ and $c'$ are called the direct effect
and total effect, respectively, of $X$ on $Y$.  The product of coefficients
$ab$ is called the indirect effect of $X$ on $Y$ and constitutes the main
parameter of interest in mediation analysis.  Under the usual assumption of
independent and normally distributed error terms $e_{1}$, $e_{2}$, and $e_{3}$,
it holds that $c' = ab + c$ \citep[e.g.,][]{mackinnon95}, and the same holds
for the corresponding ordinary least-squares (OLS) estimates.

The indirect effect $ab$ can be interpreted in the following way: a change of
one unit in $X$ explains a change of $a$ units in $M$, which in turn explains a
change of $ab$ units in $Y$.  It is therefore an important question whether or
not to standardize the variables in some way.  If the scales of the variables
differ by orders of magnitude, certain coefficients may dominate the
relationship $c' = ab + c$.  However, variables used in mediation analysis
often measure constructs that are aggregated from several rating-scale items
(e.g., on a scale of 1--5).  In such cases, a researcher may prefer not to
standardize to keep the interpretation in terms of the original measurement
scales.  Similarly, a researcher may prefer not to standardize a binary $X$
variable to keep the interpretation in terms of a change from one group to the
other.  For a more detailed discussion on whether or not to use standardized
coefficients in mediation analysis, we refer to \citet[p.~519]{hayes18}.

Mediation analysis goes back to \citet{judd81} and \citet{baron86}, however
their stepwise approach has been superseded by approaches that focus on the
indirect effect. \citet{sobel82} proposed a test for the indirect effect
that assumes a normal distribution of the corresponding estimator, which is
an unrealistic assumption for a product of coefficients.  \citet{bollen90},
\citet{shrout02}, \citet{mackinnon04}, and \citet{preacher04, preacher08}
therefore advocate for a bootstrap test based on OLS regressions, which is
the most frequently applied method for mediation analysis according to
literature reviews of \citet{wood08} and \citet{alfons22a}.  More recently,
several authors have emphasized that outliers or deviations from normality
assumptions are detrimental to the reliability and validity of mediation
analysis, and introduced more robust procedures.  \citet{zu10} propose a
bootstrap test after an initial data cleaning step, whereas \citet{yuan14}
suggest a bootstrap test based on median regressions. \citet{alfons22a}
combine the robust MM-estimator of regression \citep{yohai87} with the the
fast-and-robust bootstrap \citep{salibian02, salibian08}, and demonstrate that
this procedure outperforms the aforementioned approaches for a wide range of
error distributions (with different levels of skewness and kurtosis) and
outlier configurations.

Various software packages are available for mediation analysis.  The macro
\code{INDIRECT} \citep{preacher04, preacher08} and its successor \code{PROCESS}
\citep{hayes18} for \proglang{SPSS} \citep{SPSS} and \proglang{SAS} \citep{SAS}
implement the bootstrap test based on OLS regressions.  For the statistical
computing environment \proglang{R} \citep{R}, the general purpose packages
\pkg{psych} \citep{psych} and \pkg{MBESS} \citep{MBESS} for statistical
analysis in the behavioral sciences also provide functionality for a bootstrap
test in mediation analysis.  Package \pkg{WRS2} \citep{mair20} is a collection
of robust statistical methods, which offers mediation analysis via the
bootstrap test after data cleaning proposed by \citet{zu10}.  Other packages
concentrate on mediation analysis or specific aspects thereof.  Package
\pkg{mediation} \citep{tingley14} is focused on causal mediation analysis
in a potential outcome framework, and package \pkg{medflex} \citep{steen17}
implements recent developments in mediation analysis embedded within the
counterfactual framework.  Bayesian multilevel mediation models can be
estimated with package \pkg{bmlm} \citep{bmlm}, while package \pkg{mma}
\citep{mma} offers functionality for general multiple mediation analysis
with continuous or binary/categorical variables.  In addition, general
purpose software for structural equation modeling such as \proglang{Mplus}
\citep{Mplus} or the \proglang{R} packages \pkg{sem} \citep{sem} and
\pkg{lavaan} \citep{rosseel12} can be used for mediation analysis.  The
former also allows for maximum likelihood estimation with skew-normal,
$t$, or skew-$t$ error distributions \citep{asparouhov16}.

Despite the growing number of \proglang{R} packages that address mediation
analysis, there are no common interfaces or class structures.  Instead, each
package uses its own way of specifying mediation models and storing the
results.  Additionally, only package \pkg{WRS2} contains some functionality
for robust mediation analysis.

Package \pkg{robmed} \citep{robmed} aims to address these issues.  Its main
functionality is the robust bootstrap procedure proposed in \citet{alfons22a},
which is highly robust to outliers and other deviations from normality
assumptions.  Furthermore, \pkg{robmed} implements various other methods of
estimating mediation models, as well as different tests for the indirect
effects.  All implemented methods share the same function interface and a
clear class structure.  In addition, \pkg{robmed} introduces a simple formula
interface for specifying mediation models, and provides several plots for
diagnostics or visualization of the results from mediation analysis.

Package \pkg{robmed} is available on CRAN (the Comprehensive \proglang{R}
Archive Network, \url{https://CRAN.R-project.org/}) and can be installed
from the \proglang{R} console with the following command:
\begin{Schunk}
\begin{Sinput}
R> install.packages("robmed")
\end{Sinput}
\end{Schunk}

The rest of the paper is organized as follows.  Section~\ref{sec:methodology}
discusses various extensions of the simple mediation model, as well as the
implemented methodology for estimation and testing.
Implementation details are provided in Section~\ref{sec:implementation}, while
Section~\ref{sec:illustrations} illustrates the use of package \pkg{robmed}
with code examples. The final Section~\ref{sec:summary} concludes.




\section{Methodology} \label{sec:methodology}

We first provide overviews of the mediation models and estimation techniques
supported by package \pkg{robmed} in Sections~\ref{sec:models}
and~\ref{sec:methods}, respectively.  Section~\ref{sec:ROBMED} then gives
technical details of the robust bootstrap procedure of \citet{alfons22a}.

\subsection{Extensions of the simple mediation model} \label{sec:models}

The simple mediation model \eqref{eq:simpleM}--\eqref{eq:simpleY'} can easily
be extended in various ways, for instance with (i) multiple parallel mediators,
(ii) multiple serial mediators, and (iii) multiple independent variables to be
mediated.  All those extensions may include additional control variables
(covariates) as well.

\subsubsection{Parallel multiple mediator model}

\begin{figure}[b]
\begin{center}
\includegraphics[width=0.6\textwidth]{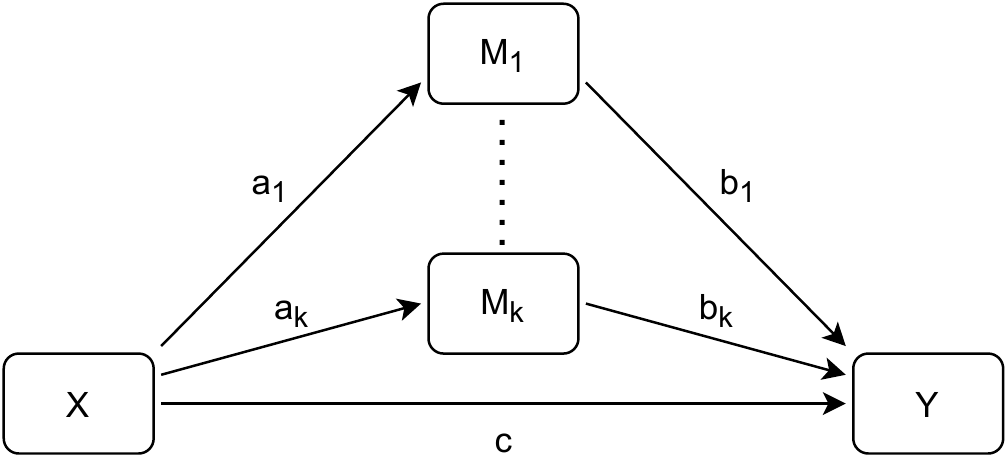}
\end{center}
\caption{Diagram visualizing a parallel multiple mediator model.}
\label{fig:parallel}
\end{figure}

In the parallel multiple mediator model, an independent variable $X$ is
hypothesized to influence a dependent variable $Y$ through multiple mediators
$M_{1}, \dots, M_{k}$, while the mediator variables do not influence each
other.  A diagram of the model is displayed in Figure~\ref{fig:parallel}, and
the corresponding regression equations are
\begin{align}
M_{j} &= i_{j} + a_{j} X + e_{j},
\qquad j = 1, \dots, k, \label{eq:parallelM} \\
Y &= i_{k+1} + b_{1} M_{1} + \dots + b_{k} M_{k} + c X + e_{k+1},
\label{eq:parallelY} \\
Y &= i_{k+2} + c' X + e_{k+2}, \label{eq:parallelY'}
\end{align}
where $i_{1}, \dots, i_{k+2}$, $a_{1}, \dots, a_{k}$, $b_{1}, \dots, b_{k}$,
$c$, and $c'$ are regression coefficients to be estimated, and $e_{1}, \dots,
e_{k+2}$ are random error terms.  With the usual assumptions of independent
and normally distributed error terms, we now have that $c' = \sum_{j = 1}^{k}
a_{j} b_{j} + c$.  The main parameters of interest are the individual indirect
effects $a_{1}b_{1}, \dots, a_{k}b_{k}$, and it can also be of interest to make
pairwise comparisons between the individual indirect effects or their absolute
values \citep[e.g.,][Chapter~5.1]{hayes18} if the hypothesized mediators are
scaled similarly.

\subsubsection{Serial multiple mediator model}

\begin{figure}[t]
\begin{center}
\includegraphics[width=0.7\textwidth]{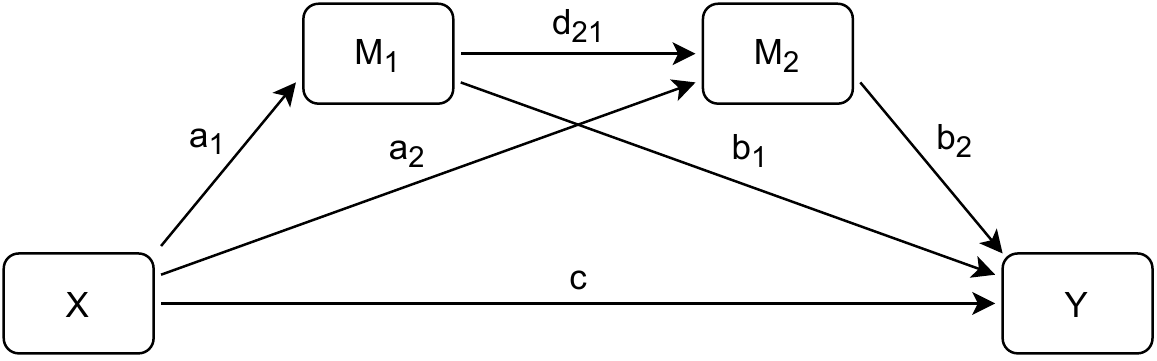}
\end{center}
\caption{Diagram visualizing a serial multiple mediator model with two
mediators.}
\label{fig:serial}
\end{figure}

A distinctive feature of the serial multiple mediator model is that the
hypothesized mediators $M_{1}, \dots, M_{k}$ may influence each other in a
sequential manner, unlike the parallel multiple mediator model in which the
mediators do not affect one another.  Figure~\ref{fig:serial} contains a
diagram of the model with two serial mediators, while the model in its general
form is given by
\begin{align}
\begin{split}
M_{1} &= i_{1} + a_{1} X + e_{1}, \\
M_{2} &= i_{2} + d_{21} M_{1} + a_{2} X + e_{2}, \\
      &\hspace{0.5em} \vdots \\
M_{k} &= i_{k} + d_{k1} M_{1} + \dots + d_{k,k-1} M_{k-1} + a_{k} X + e_{k},
\end{split} \label{eq:serialM} \\
Y &= i_{k+1} + b_{1} M_{1} + \dots + b_{k} M_{k} + c X + e_{k+1},
\label{eq:serialY} \\
Y &= i_{k+2} + c' X + e_{k+2}, \label{eq:serialY'}
\end{align}
where $i_{1}, \dots, i_{k+2}$, $a_{1}, \dots, a_{k}$, $d_{j1}, \dots,
d_{j,j-1}$, $j = 2, \dots, k$, $b_{1}, \dots, b_{k}$, $c$, and $c'$ are
regression coefficients to be estimated, and $e_{1}, \dots, e_{k+2}$ are
random error terms.  It is easy to see that the serial multiple mediator model
quickly grows in complexity with increasing number of mediators due to the
combinatorial increase in indirect paths through the mediators (the number of
indirect paths is given by $\sum_{j = 1}^{k} {k \choose j}$ for $k$ serial
mediators).  In package \pkg{robmed}, it is therefore only implemented for
two and three mediators to maintain a focus on easily interpretable models.
Here, we only discuss the model for two serial mediators, and we refer to
\citet[p.169--171]{hayes18} for a diagram and a description of the various
indirect effects in the case of three serial mediators.

For two serial mediators, the three indirect effects
$a_{1}b_{1}$ ($X \rightarrow M_{1} \rightarrow Y$),
$a_{2}b_{2}$ ($X \rightarrow M_{2} \rightarrow Y$), and
$a_{1}d_{21}b_{2}$ ($X \rightarrow M_{1} \rightarrow M_{2} \rightarrow Y$)
are the main parameters of interest.  However, not all pairwise comparisons
of the indirect effects may be meaningful (even if the mediators are scaled
similarly), as $a_{1}d_{21}b_{2}$ can be expected to be different in magnitude
from $a_{1}b_{1}$ and $a_{2}b_{2}$.  Finally, we have that $c' = a_{1}b_{1} +
a_{2}b_{2} + a_{1}d_{21}b_{2} + c$ under the usual assumptions of independent
and normally distributed error terms.

\subsubsection{Multiple independent variables to be mediated}

\begin{figure}[t!]
\begin{center}
\includegraphics[width=0.525\textwidth]{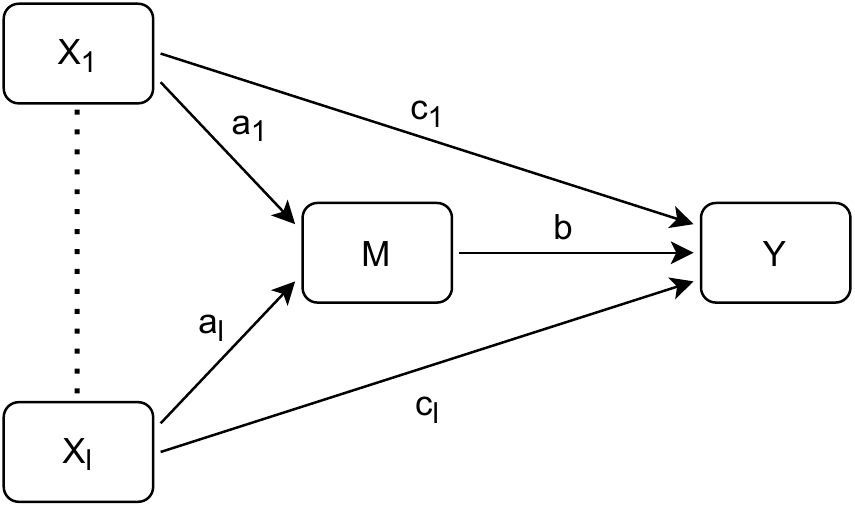}
\end{center}
\caption{Diagram visualizing a mediation model with multiple independent
variables.}
\label{fig:multiple}
\end{figure}

Instead of having multiple mediators, one can also allow for multiple
independent variables $X_{1}, \dots, X_{l}$ to influence the dependent
variable $Y$ through a hypothesized mediator $M$.  The resulting model
is visualized in Figure~\ref{fig:multiple} and defined by the equations
\begin{align}
M &= i_{1} + a_{1} X_{1} + \dots + a_{l} X_{l} + e_{1}, \label{eq:multipleM} \\
Y &= i_{2} + b M + c_{1} X_{1} + \dots + c_{l} X_{l} + e_{2},
\label{eq:multipleY} \\
Y &= i_{3} + c_{1}' X_{1} + \dots + c_{l}' X_{l} + e_{3}, \label{eq:parallelY'}
\end{align}
where $i_{1}, i_{2}, i_{3}$, $a_{1}, \dots, a_{l}$, $b$, $c_{1}, \dots, c_{l}$,
and $c_{1}', \dots, c_{l}'$ are regression coefficients to be estimated, and
$e_{1}$, $e_{2}$, and $e_{3}$ are random error terms.  The indirect effects
$a_{1}b, \dots, a_{l}b$ are the main parameters of interest, and with the
direct effects $c_{1}, \dots, c_{l}$ and total effects $c_{1}', \dots, c_{l}'$,
it holds that $c_{j}' = a_{j}b + c_{j}$, $j = 1, \dots, l$, under the usual
independence and normality assumptions on the error terms. If the independent
variables are on a comparable scale, it can also be of interest to make
pairwise comparisons between the indirect effects or their absolute values.

This model is commonly used when the hypothesized mediator is the main
(explanatory) variable of interest and its antecedents are being studied.
Furthermore, an important special case of this model occurs when a categorical
independent variable is represented by a group of dummy variables.

\subsubsection{Control variables}

In many study designs, it may be necessary to isolate the effects of the
independent variables of interest from other factors.  For instance, consider
a study on whether exercise-induced feelings such as physical exhaustion
mediate the relationship between physical activity and depression
\citep[cf.][]{pickett12}.  If the participants vary in demographics such as
age and gender, the researcher may need to control for the effects of those
variables \citep[e.g., older people may be less capable of engaging in
strenuous activities;][]{cerin09}.  Such control variables should be added to
all regression equations of a mediation model.  This means that there is no
intrinsic difference between independent variables of interest and control
variables in terms of the model or its estimation.  The difference is purely
conceptual in nature: for the control variables, the estimates of the direct
and indirect paths are not of particular interest to the researcher.  Package
\pkg{robmed} therefore allows to specify control variables separately from the
independent variables of interest.  Only for the latter, results for the
indirect effects are included in the output.

While we omitted control variables from the above equations and diagrams for
notational simplicity, package \pkg{robmed} supports additional control
variables in all implemented models.

\subsubsection{More complex models}

The models described above do not exist in isolation and some of them can be
combined.  For instance, \pkg{robmed} supports parallel and serial multiple
mediator models with multiple independent variables of interest.  Other
variations of the mediation model, such as the moderated mediation and
mediated moderation models \citep[e.g.,][]{muller05} are out of scope for
this paper. They are not yet implemented in package \pkg{robmed} but we aim
to add support in future versions.

\subsection{Overview of implemented methods} \label{sec:methods}

\begin{sidewaystable}
\centering
\begin{tabular}{llp{1.6cm}lp{8.75cm}ccccc}
\hline\noalign{\smallskip}
\code{test} & \code{method} & \code{robust} & \code{family} & Description &
\rotatebox{90}{Simple mediation} &
\rotatebox{90}{Parallel mediators} &
\rotatebox{90}{Serial mediators} &
\rotatebox{90}{\makecell[l]{Multiple independent \\ variables of interest}} &
\rotatebox{90}{Control variables} \\
\noalign{\smallskip}\hline\noalign{\medskip}
\code{"boot"} & \code{"regression"} & \code{"MM"} or $\enskip$ \code{TRUE} & &
MM-regression \citep{yohai87} and the fast-and-robust bootstrap
\citep{salibian02} are used to construct a confidence interval
\citep{alfons22a}. &
\checkmark & \checkmark & \checkmark & \checkmark & \checkmark \\
\noalign{\smallskip}
\code{"boot"} & \code{"regression"} & \code{"median"} & &
A bootstrap confidence interval is computed based on median regressions
\citep{yuan14}. &
\checkmark & \checkmark & \checkmark & \checkmark & \checkmark \\
\noalign{\smallskip}
\code{"boot"} & \code{"regression"} & \code{FALSE} & \code{"gaussian"} &
A bootstrap confidence interval is computed based on OLS regressions
\citep{bollen90, shrout02, mackinnon04, preacher04, preacher08}. &
\checkmark & \checkmark & \checkmark & \checkmark & \checkmark \\
\noalign{\smallskip}
\code{"boot"} & \code{"regression"} & \code{FALSE} & \code{"select"} &
Regression models with normal, skew-normal, $t$, or skew-$t$ error
distributions are estimated \citep{azzalini13}, and a bootstrap confidence
interval is computed. The best fitting error distribution is selected via
BIC. &
\checkmark & \checkmark & \checkmark & \checkmark & \checkmark \\
\noalign{\smallskip}
\code{"boot"} & \code{"covariance"} & \code{TRUE} & &
Following multivariate winsorization, a bootstrap confidence interval is
computed for coefficient \mbox{estimation} via the maximum likelihood estimator
of the covariance matrix \citep{zu10}. &
\checkmark & & & & \\
\noalign{\smallskip}
\code{"boot"} & \code{"covariance"} & \code{FALSE} & &
A bootstrap confidence interval is computed for coefficient estimation
via the maximum likelihood \mbox{estimator} of the covariance matrix. &
\checkmark & & & & \\
\noalign{\smallskip}
\noalign{\medskip}\hline\noalign{\bigskip}
\end{tabular}
\caption{Overview of bootstrap procedures for mediation analysis implemented in
the \proglang{R} package \pkg{robmed}, and corresponding argument values to use
in function \code{test\_mediation()}.}
\label{tab:methods-boot}
\end{sidewaystable}

\begin{sidewaystable}
\centering
\begin{tabular}{llp{1.6cm}lp{8.5cm}ccccc}
\hline\noalign{\smallskip}
\code{test} & \code{method} & \code{robust} & \code{family} & Description &
\rotatebox{90}{Simple mediation} &
\rotatebox{90}{Parallel mediators} &
\rotatebox{90}{Serial mediators} &
\rotatebox{90}{\makecell[l]{Multiple independent \\ variables of interest}} &
\rotatebox{90}{Control variables} \\
\noalign{\smallskip}\hline\noalign{\medskip}
\code{"sobel"} & \code{"regression"} & \code{"MM"} or $\enskip$ \code{TRUE} & &
A variation of the Sobel test based on coefficient estimation via
MM-regressions. &
\checkmark & & & & \checkmark \\
\code{"sobel"} & \code{"regression"} & \code{"median"} & &
A variation of the Sobel test based on coefficient estimation via median
regressions. &
\checkmark & & & & \checkmark \\
\noalign{\medskip}
\code{"sobel"} & \code{"regression"} & \code{FALSE} & \code{"gaussian"} &
A variation of the Sobel test based on coefficient estimation via OLS
regressions. &
\checkmark & & & & \checkmark \\
\noalign{\medskip}
\code{"sobel"} & \code{"regression"} & \code{FALSE} & \code{"select"} &
A variation of the Sobel test in which regression models with normal,
skew-normal, $t$, or skew-$t$ error distributions are estimated, with
the best fitting distribution being selected via BIC. &
\checkmark & & & & \checkmark \\
\noalign{\medskip}
\code{"sobel"} & \code{"covariance"} & \code{TRUE} & &
Following multivariate winsorization, a variation of the Sobel test is applied
based on coefficient \mbox{estimation} via the maximum likelihood estimator of
the covariance matrix. &
\checkmark & & & & \\
\code{"sobel"} & \code{"covariance"} & \code{FALSE} & &
A variation of the Sobel test based on coefficient \mbox{estimation} via the
maximum likelihood estimator of the covariance matrix. &
\checkmark & & & & \\
\noalign{\medskip}
\noalign{\medskip}\hline\noalign{\bigskip}
\end{tabular}
\caption{Overview of variations of the Sobel test \citep{sobel82} for
mediation analysis implemented in the \proglang{R} package \pkg{robmed}, and
corresponding argument values to use in function \code{test\_mediation()}.
Those tests are included for benchmarking purposes and are not recommended for
empirical analyses.}
\label{tab:methods-sobel}
\end{sidewaystable}

While package \pkg{robmed} is focused on the fast-and-robust bootstrap
procedure for mediation analysis introduced by \citet{alfons22a}, various
other methods are available as well.  Tables~\ref{tab:methods-boot}
and~\ref{tab:methods-sobel} provide an overview of the available methods
together with the corresponding argument values to use in function
\code{test_mediation()}, which implements mediation analysis in
\pkg{robmed}.

A bootstrap test is considered state-of-the art for mediation analysis, with
many authors advocating to use the bootstrap with ordinary least-squares (OLS)
estimation of the coefficients in the mediation model \citep{bollen90,
shrout02, mackinnon04, preacher04, preacher08}.
However, a bootstrap test can easily be applied to other methods of estimation.
For instance, the mediation model can also be estimated via regressions with
more flexible error distributions such as the skew-normal, $t$, or skew-$t$
distributions \citep[see][for maximum likelihood estimation of such regression
models]{azzalini13}.
Note that a similar procedure for mediation analysis, but using structural
equation modeling, has been suggested in \citet{asparouhov16}.
Package \pkg{robmed} goes a step further in that it allows to select the best
fitting error distribution via the Bayesian information criterion (BIC)
\citep{schwarz78}.
In addition, other robust methods for mediation analysis are implemented in
\pkg{robmed}.
\citet{yuan14} proposed a bootstrap test that replaces OLS estimation with
median regression.
\citet{zu10} proposed to first winsorize the data via a Huber M-estimator of
the covariance matrix, and then to perform a bootstrap test on the cleaned data
with coefficient estimation based on the maximum likelihood covariance matrix.
A discussion of advantages and disadvantages of those approaches, as well as a
comparison in extensive simulation studies, can be found in \citet{alfons22a}.

Besides bootstrap tests, variations of the Sobel test \citep{sobel82} are
implemented in \pkg{robmed}.
The Sobel test was originally proposed for maximum likelihood estimation of
structural equation models, of which mediation models are a special case.
It assumes a normal distribution of the indirect effect estimator and
simplifies the calculation of the standard error by taking a first- or
second-order approximation.
This test has been criticized in the literature for its incorrect assumptions
\citep[e.g.,][]{mackinnon02}, and a bootstrap test is generally preferred.
Nevertheless, the Sobel test can easily be generalized to other
\mbox{estimation} methods, and it is implemented in \pkg{robmed} for all
estimation procedures of the (simple) mediation model.  We emphasize that the
Sobel tests are implemented for comparisons in benchmarking experiments and
that they are not recommended for empirical analyses.

\subsection{Fast-and-robust bootstrap test for mediation analysis}
\label{sec:ROBMED}

The robust procedure of \citet{alfons22a} follows the state-of-the-art
bootstrap approach for testing mediation \citep{bollen90, shrout02,
mackinnon04, preacher04, preacher08}, but it replaces OLS regressions
with the robust MM-estimator of regression \citep{yohai87} and the standard
bootstrap with the fast-and-robust bootstrap \citep{salibian02, salibian08}.

\subsubsection{Robust regression}

For a response variable $Y$, a $(p+1)$-dimensional random vector $\vect{X}$
in which the first component is fixed at~1, and a random error term
$\varepsilon \sim N(0, \sigma^{2})$, the linear regression model is given by
\begin{equation*}
Y = \vect{X}^\top \vect{\beta} + \varepsilon.
\end{equation*}
Denoting the corresponding observations by $(y_{i}, \obs{x}_{i}^\top)^\top$,
$i = 1, \dots, n$, the MM-estimate of regression with loss function $\rho$
\citep{yohai87} is defined as
\begin{equation} \label{eq:MM}
\vect{\hat{\beta}}_{n} = \argmin_{\beta} \sum_{i=1}^{n}
\rho \left( \frac{r_{i}(\vect{\beta})}{\hat{\sigma}_{n}} \right),
\end{equation}
where $r_{i}(\vect{\beta}) = y_{i} - \obs{x}_{i}^\top \vect{\beta}$,
$i = 1, \dots, n$, are the residuals, and $\hat{\sigma}_{n}$ is an initial
estimate of the residual scale.  We take $\hat{\sigma}_{n}$ from a highly
robust but inefficient S-estimator of regression \citep{rousseeuw84,
salibian06}, i.e.,
\begin{equation*}
\hat{\sigma}_{n} = \min_{\vect{\beta}} \hat{\sigma}_{n}(\vect{\beta}),
\end{equation*}
where $\hat{\sigma}_{n}(\vect{\beta})$ is defined implicitly as the solution of
\begin{equation*}
\frac{1}{n} \sum_{i=1}^{n} \rho_{S}
\left( \frac{r_{i}(\vect{\beta})}{\hat{\sigma}_{n}(\vect{\beta})} \right) =
\delta
\end{equation*}
with loss function $\rho_{S}$ and $\delta = E \left[ \rho_{S} \left(
\frac{X}{\sigma} \right) \right]$ for 
a random variable $X \sim N(0, \sigma^{2})$.
For both $\rho$ and $\rho_{S}$, we use Tukey's bisquare loss function defined
as
\begin{equation} \label{eq:loss}
\rho(x) = \left\{
\begin{array}{ll}
\displaystyle
\frac{x^{6}}{6c^{4}} - \frac{x^4}{2c^{2}} + \frac{x^{2}}{2},
 & \text{if } |x| \leq c \\
\noalign{\smallskip}
\displaystyle
\frac{c^{2}}{6},
 & \text{if } |x| > c.
\end{array}
\right.
\end{equation}
The value of the tuning constant $c$ in $\rho_{S}$ determines the robustness
of the MM-estimator, and the value of $c$ in $\rho$ determines the efficiency
\citep[cf.][]{yohai87}.  By default, we set $c = 1.54764$ in $\rho_{S}$ for
maximal robustness and $c = 3.443689$ in $\rho$ for 85\% efficiency at the
model with normally distributed errors.

\begin{figure}[t!]
\begin{center}
\includegraphics[width=0.975\textwidth]{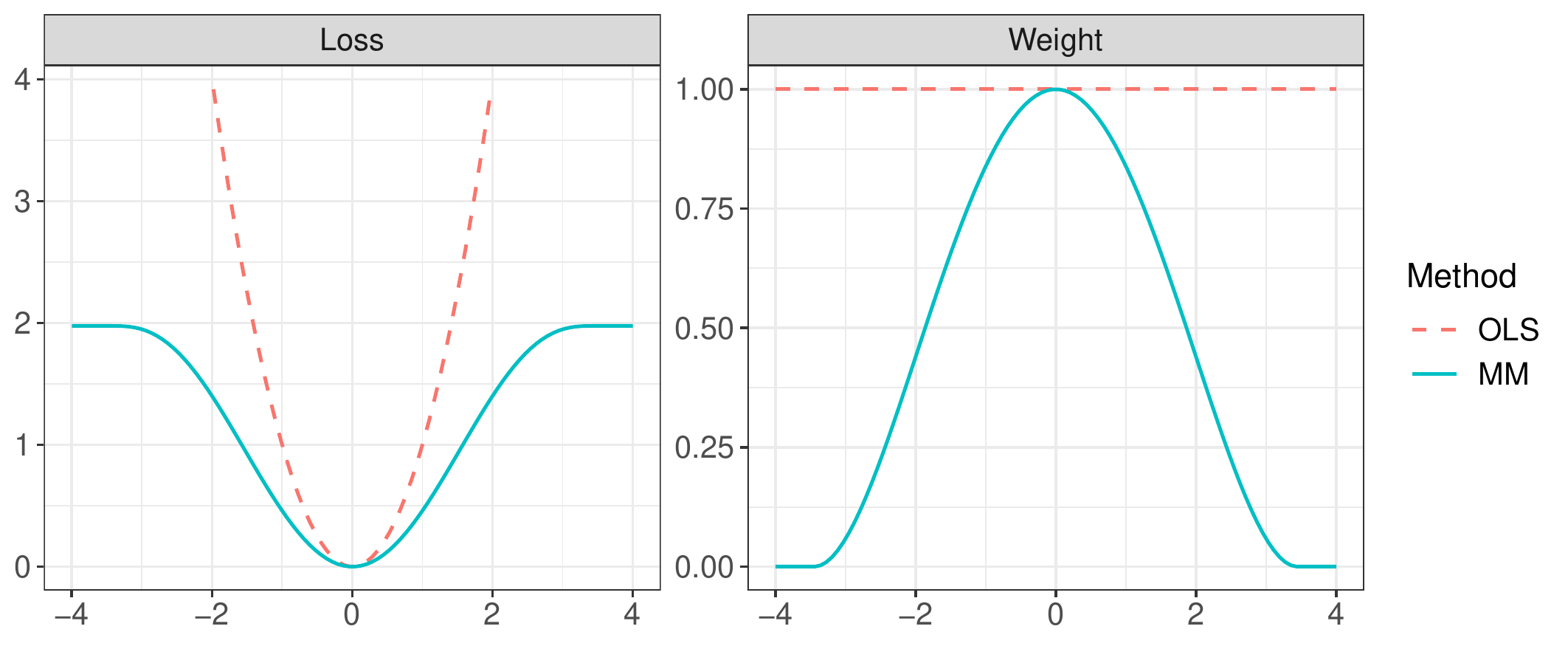}
\caption{Loss function (left) and corresponding weight function (right) for OLS
regression and the robust MM-estimator of regression.}
\label{fig:loss}
\end{center}
\end{figure}

Taking the derivative of the objective function in~\eqref{eq:MM} and equating
the derivative to~$\vect{0}$ yields the system of equations
\begin{equation} \label{eq:EstEq}
\sum_{i=1}^{n} \rho' \left( \frac{r_{i}(\vect{\beta})}{\hat{\sigma}_{n}} \right)
\obs{x}_{i} = \vect{0}.
\end{equation}
With weights
\begin{equation*}
w_{i} =
\frac{\rho'(r_{i}(\vect{\beta})/\hat{\sigma}_{n})}%
{r_{i}(\vect{\beta})/\hat{\sigma}_{n}},
\qquad i = 1, \dots, n,
\end{equation*}
the system of equations in \eqref{eq:EstEq} can be rewritten as a
weighted version of the normal equations:
\begin{equation} \label{eq:wEstEq}
\sum_{i=1}^{n} w_{i} r_{i}(\vect{\beta}) \obs{x}_{i} = \vect{0}.
\end{equation}
Therefore, the MM-estimator can be seen as a weighted least-squares estimator
with data-dependent weights. Those weights indicate how much each observation
deviates, as an observation with a large residual (large deviation) receives a
weight of 0 or close to 0, while an observation with a small residual (small
deviation) receives a weight close to 1. The loss function from \eqref{eq:loss}
and the resulting weight function are displayed in Figure~\ref{fig:loss}, which
also includes the loss function and weight function from OLS regression for
comparison.

\subsubsection{Fast-and-robust bootstrap}

Here we briefly present the main idea of the fast-and-robust bootstrap.  For
a detailed discussion and complete derivations, the reader is referred to
\citet{salibian02} and \citet{salibian08}.  There are two concerns with
bootstrapping robust estimators:
\begin{enumerate}
  \item Outliers may be oversampled in some bootstrap samples to the extent
  that those samples contain more outliers than the robust estimator can
  handle, in which case bootstrap confidence intervals become unreliable.
  \item Robust estimators typically have higher computational complexity
  than their nonrobust counterparts, which is amplified when computing many
  bootstrap replicates.
\end{enumerate}
In many empirical applications of mediation analysis, the first concern is
unlikely to be an issue when using the MM-estimator of regression due to its
high robustness.  We therefore use the fast-and-robust bootstrap mainly for
its computational efficiency, although the extra robustness does  provide
additional peace of mind.

To derive the fast-and-robust bootstrap for the MM-estimator, note that the
solution of \eqref{eq:wEstEq} can be written as
\begin{equation*}
\vect{\hat{\beta}}_{n} =
\left( \sum_{i=1}^{n} w_{i} \obs{x}_{i} \obs{x}_{i}^\top \right)^{-1}
\sum_{i=1}^{n} w_{i} \obs{x}_{i} y_{i}.
\end{equation*}
For a bootstrap sample $(y_{i}^{*}, {\obs{x}_{i}^{*}}^\top)^\top$, $i = 1,
\dots, n$, one can compute $r_{i}^{*} = y_{i}^{*} - {\obs{x}_{i}^{*}}^\top
\vect{\hat{\beta}}_{n}$ and $w_{i}^{*} = \rho'(r_{i}^{*}/\hat{\sigma}_{n}) /
(r_{i}^{*}/\hat{\sigma}_{n})$ for $i = 1, \dots, n$.
It is important to note that $\vect{\hat{\beta}}_{n}$ and $\hat{\sigma}_{n}$
are computed in advance on the original sample such that the robustness
weights $w_{i}^{*}$ are inherited from the respective observations in the
original sample.  Then only a weighted least-squares fit is computed on the
bootstrap sample to obtain
\begin{equation} \label{eq:WLS}
\vect{\hat{\beta}}_{n}^{\text{WLS}} =
\left( \sum_{i=1}^{n} w_{i}^{*} \obs{x}_{i}^{*} {\obs{x}_{i}^{*}}^\top
\right)^{-1} \sum_{i=1}^{n} w_{i}^{*} \obs{x}_{i}^{*} y_{i}^{*}.
\end{equation}
However, using \eqref{eq:WLS} for the bootstrap distribution would not capture
all the variability in the MM-estimator, as the robustness weights are not
recomputed on the bootstrap samples.  Nevertheless, a linear correction of
the coefficients can be applied to overcome this loss of variability.  The
correction matrix only needs to be computed once based on the original sample
and is given by
\begin{equation*}
\mat{K}_{n} =
\left( \sum_{i=1}^{n} \rho''(r_{i}/\hat{\sigma}_{n}) \obs{x}_{i}
\obs{x}_{i}^\top \right)^{-1} \sum_{i=1}^{n} w_{i} \obs{x}_{i}
\obs{x}_{i}^\top.
\end{equation*}
Then the fast-and-robust bootstrap replicate on the given bootstrap sample is
computed as
\begin{equation} \label{eq:FRB}
\vect{\hat{\beta}}_{n}^{*} = \vect{\hat{\beta}}_{n} + \mat{K}_{n}
\left( \vect{\hat{\beta}}_{n}^{\text{WLS}} - \vect{\hat{\beta}}_{n} \right).
\end{equation}
Since the MM-estimator $\vect{\hat{\beta}}_{n}$ is consistent for
$\vect{\beta}$ \citep{yohai87}, $\sqrt{n} (\vect{\hat{\beta}}_{n}^{*} -
\vect{\hat{\beta}}_{n})$ has the same asymptotic distribution as $\sqrt{n}
(\vect{\hat{\beta}}_{n} - \vect{\beta})$ \citep{salibian08, salibian02}.

\subsubsection{Bootstrapping the indirect effects in mediation analysis}

For simplicity, we focus on the indirect effect in the simple mediation model
from~\eqref{eq:simpleM}--\eqref{eq:simpleY'}. Similar calculations apply to the
indirect effects in the mediation models described in Section~\ref{sec:models}.
On each bootstrap sample, \eqref{eq:FRB} is used to obtain estimates $\hat{a}$,
$\hat{b}$, and $\hat{c}$ of the coefficients in~\eqref{eq:simpleM}
and~\eqref{eq:simpleY}, and therefore estimates $\hat{a}\hat{b}$ of the
indirect effect.  Note that we do not perform the regression corresponding
to~\eqref{eq:simpleY'} and instead estimate the total effect by $\hat{c}' =
\hat{a}\hat{b} + \hat{c}$.  With the empirical distribution of the indirect
effect over the bootstrap samples, we construct a percentile-based confidence
interval.  By default, we report a bias-corrected and accelerated confidence
interval \citep{davison97}.  Furthermore, we advocate to report the mean over
the bootstrap distribution as the final point estimates of the indirect effect.


\section{Package contents and implementation} \label{sec:implementation}

We describe the included data set in Section~\ref{sec:data}, introduce the
formula interface for specifying mediation models in Section~\ref{sec:formula},
and briefly discuss the main functions as well as the class structure of
package \pkg{robmed} in Section~\ref{sec:classes}.  Moreover, we load the
package and the data in order to use them in code examples.
\begin{Schunk}
\begin{Sinput}
R> library("robmed")
R> data("BSG2014")
\end{Sinput}
\end{Schunk}

\subsection{Example data} \label{sec:data}

The \code{BSG2014} data included in package \pkg{robmed} come from a business
simulation game that was played by senior business administration students as
part of a course at a Western European university.  The simulation game was
played twice, and a survey was conducted in three waves (before the first game,
in between the two games, and after the second game).  A total of 354 students
formed 92 randomly assigned teams, and the responses of the individual students
were aggregated to the team level. Leaving out teams with less than 50 percent
response rate yields a sample size of $n = 89$ teams.

Below, we provide an overview of the variables that are used later on in the
case studies in Section~\ref{sec:illustrations}.  For a complete description
of the data, we refer to its \proglang{R} help file, which can be accessed
from the console with \code{?BSG2014}.

\begin{description}
  \item{\code{ValueDiversity}:}
  Using the short Schwartz’s value survey \citep{lindeman05}, the team members
  rated ten items on the importance of certain values (1~= not important, 10 =
  highly important).  For each value item, the coefficient of variation of the
  individual responses across team members was computed, and the resulting
  coefficients of variation were averaged across the value items.
  \item{\code{TaskConflict}:} Using the intra-group conflict scale of
  \citet{jehn95}, the team members rated four items on the presence of conflict
  regarding the work on a 5-point scale (1~= none, 5~= a lot).  The individual
  responses were aggregated by taking the average across items and team
  members.
  \item{\code{TeamCommitment}:} The team members indicated the extent to which
  they agree or disagree with four items on commitment to the team, which are
  based on \citet{mowday79}, using a 5-point scale (1~= strongly disagree,
  5~= strongly agree).  The individual responses were aggregated by taking the
  average across items and team members.
  \item{\code{TeamScore}:} The team performance scores on the second game were
  computed at the end of the simulation through a mix of five objective
  performance measures: return on equity, earnings-per-share, stock price,
  credit rating, and image rating. The computation of the scores is handled by
  the simulation game software, and details can be found in \citet{mathieu09}.
  \item{\code{SharedLeadership}:} Following \citet{carson07}, every team member
  assessed each of their peers on the question of ``To what degree does your
  team rely on this individual for leadership?'' using a 5-point scale (1~= not
  at all, 5~= to a very large extent).  The leadership ratings were aggregated
  by taking the sum and dividing it by the number of pairwise relationships
  among team members.
  \item{\code{AgeDiversity}:} Following \citet{harrison07}, age diversity was
  operationalized by the coefficient of variation of the team members' ages.
  \item{\code{GenderDiversity}:} Gender diversity was measured with Blau's
  index, $1 - \sum_{j} p_{j}^{2}$, where $p_{j}$ is the proportion of team
  members in the $j$-th category \citep{blau77}.
  \item{\code{ProceduralJustice}:} Based on the intra-unit procedural justice
  climate scale of \citet{li09}, the team members indicated the extent to which
  they agree or disagree with four items on a 5-point scale (1~= strongly
  disagree, 5~= strongly agree).  The individual responses were aggregated by
  taking the average across items and team members.
  \item{\code{InteractionalJustice}:} Using the intra-unit interactional
  justice climate scale of \citet{li09}, the team members indicated the
  extent to which they agree or disagree with four items on a 5-point scale
  (1~= strongly disagree, 5~= strongly agree).  The individual responses were
  aggregated by taking the average across items and team members.
  \item{\code{TeamPerformance}:} Following \citet{hackman86}, the team members
  indicated the extent to which they agree or disagree with four items on the
  team's functioning, using a 5-point scale (1~= strongly disagree, 5~=
  strongly agree).  The individual responses were aggregated by taking the
  average across items and team members.
\end{description}

To gain some insight into the distribution of those variables (including their
ranges), we extract them from the data set and produce a summary:
\begin{Schunk}
\begin{Sinput}
R> keep <- c("ValueDiversity", "TaskConflict", "TeamCommitment", "TeamScore",
+            "SharedLeadership", "AgeDiversity", "GenderDiversity",
+            "ProceduralJustice", "InteractionalJustice", "TeamPerformance")
R> summary(BSG2014[, keep])
\end{Sinput}
\begin{Soutput}
 ValueDiversity   TaskConflict   TeamCommitment    TeamScore     
 Min.   :1.105   Min.   :1.125   Min.   :2.125   Min.   : 49.00  
 1st Qu.:1.443   1st Qu.:1.500   1st Qu.:3.625   1st Qu.: 90.00  
 Median :1.587   Median :1.688   Median :3.875   Median : 98.00  
 Mean   :1.676   Mean   :1.761   Mean   :3.822   Mean   : 95.72  
 3rd Qu.:1.916   3rd Qu.:2.000   3rd Qu.:4.125   3rd Qu.:104.00  
 Max.   :2.548   Max.   :2.938   Max.   :4.688   Max.   :110.00  
 SharedLeadership  AgeDiversity    GenderDiversity  ProceduralJustice
 Min.   :3.500    Min.   :0.0000   Min.   :0.0000   Min.   :3.375    
 1st Qu.:6.333    1st Qu.:0.5000   1st Qu.:0.0000   1st Qu.:3.750    
 Median :6.667    Median :0.8165   Median :0.3750   Median :3.875    
 Mean   :6.629    Mean   :0.9723   Mean   :0.3031   Mean   :3.908    
 3rd Qu.:7.167    3rd Qu.:1.2583   3rd Qu.:0.3750   3rd Qu.:4.062    
 Max.   :9.333    Max.   :4.2720   Max.   :0.5000   Max.   :4.500    
 InteractionalJustice TeamPerformance
 Min.   :3.312        Min.   :3.000  
 1st Qu.:4.167        1st Qu.:3.667  
 Median :4.375        Median :4.000  
 Mean   :4.379        Mean   :3.968  
 3rd Qu.:4.625        3rd Qu.:4.250  
 Max.   :5.000        Max.   :4.750  
\end{Soutput}
\end{Schunk}
For instance, the objective team performance scores in variable
\code{TeamScore} range from~$49$
to~$110$.

\subsection{Formula interface} \label{sec:formula}

The equations in the mediation model follow a specific structure regarding
which variable is used as the response variable and which variables are the
explanatory variables.  Some \proglang{R} packages for mediation analysis,
e.g., \pkg{mediation} \citep{tingley14}, require the user to specify one
formula for each equation, which can be tedious and prone to mistakes, in
particular for models with multiple mediators and multiple independent
variables or control variables.  Other packages, e.g., \pkg{psych}
\citep{psych} or \pkg{MBESS} \citep{MBESS}, do not provide a formula
interface at all, despite formulas being the standard way of describing
models in \proglang{R}.

For package \pkg{robmed}, we designed a formula interface that builds upon the
standard formula interface in \proglang{R}, but allows to specify the
mediation model with a single formula.  As usual, the dependent variable is
defined on the left hand side of the formula, and the independent variable is
given on the right hand side.  In addition, the functions \code{m()} and
\code{covariates()} can be used on the right hand side to define the
hypothesized mediators and any control variables, respectively.
If multiple mediators are supplied, function \code{m()} provides the argument
\code{.model}, which accepts the values \code{"parallel"} for parallel
mediators (the default) and \code{"serial"} for serial mediators.  The
corresponding wrapper functions \code{parallel_m()} and \code{serial_m()} are
available for convenience.

For example, a simple mediation model can be defined as follows (see also the
case study in Section~\ref{sec:simple}):
\begin{Schunk}
\begin{Sinput}
R> TeamCommitment ~ m(TaskConflict) + ValueDiversity
\end{Sinput}
\end{Schunk}
An example for a serial multiple mediator model is specified with the following
formula (see also the case study in Section~\ref{sec:serial}), where the serial
mediators are listed in consecutive order from left to right:
\begin{Schunk}
\begin{Sinput}
R> TeamScore ~ serial_m(TaskConflict, TeamCommitment) + ValueDiversity
\end{Sinput}
\end{Schunk}
The formula specification for an example of a parallel multiple mediator
model with control variables is given by (see also the case study in
Section~\ref{sec:parallel}):
\begin{Schunk}
\begin{Sinput}
R> TeamPerformance ~ parallel_m(ProceduralJustice, InteractionalJustice) +
+    SharedLeadership + covariates(AgeDiversity, GenderDiversity)
\end{Sinput}
\end{Schunk}
Note that different variables within \code{m()}, \code{parallel_m()},
\code{serial_m()}, and \code{covariates()} are separated by commas.

\subsection{Main functions and class structure} \label{sec:classes}

The two main functions of package \pkg{robmed} are \code{fit_mediation()},
which implements various methods for the estimation of a mediation model,
and \code{test_mediation()}, which performs statistical tests on the indirect
effects in the mediation model.  Furthermore, \pkg{robmed} follows a clear
object-oriented design using \code{S3} classes \citep{chambers92}.

Function \code{fit_mediation()} is mainly intended to be used internally by
\code{test_mediation()}, but it is also useful for a user who wants to compare
different tests on the indirect effects for the same method of estimation, such
that the estimation on the given sample only has to be performed once.
It returns an object inheriting from class \code{"fit_mediation"}.  The
currently available subclasses are \code{"reg_fit_mediation"} if the mediation
model was estimated via a series of regressions, and \code{"cov_fit_mediation"}
if the model was estimated based on the covariance matrix of the involved
variables.

We expect most users to find it more convenient to use \code{test_mediation()}
directly in order to perform model estimation and testing the indirect effects
with one function call.  See Tables~\ref{tab:methods-boot}
and~\ref{tab:methods-sobel} for an overview of which argument values to use
in \code{test_mediation()} for the various available methods.  Furthermore,
function \code{robmed()} is a wrapper function for the fast-and-robust
bootstrap test of \citet{alfons22a}.  The results are returned as an object
inheriting from class \code{"test_mediation"}.  The currently available
subclasses are \code{"boot_test_mediation"} for bootstrap tests, and
\code{"sobel_test_mediation"} for tests based on the normal approximation of
\citet{sobel82}.  Among other information, the component \code{fit} stores the
estimation results as an object inheriting from class \code{"fit_mediation"}.
Objects of class \code{"boot_test_mediation"} also contain a component
\code{reps}, which stores the bootstrap replicates as an object of class
\code{"boot"}, as returned by function \code{boot()} from package \pkg{boot}
\citep{boot}.  It should be noted that the internal use of function
\code{boot()} implies that the user can easily take advantage of parallel
computing to reduce computation time.

Functions \code{fit_mediation()} and \code{test_mediation()} are implemented
as generic functions.  Two methods are available for both functions: one method
uses the formula interface described in Section~\ref{sec:formula}, while the
default method provides an alternative way of specifying mediation models.
Additionally, \code{test_mediation()} has a method for objects inheriting from
class \code{"fit_mediation"}, as returned by \code{fit_mediation()}.  The
default methods take the data set as their first argument in order to work
nicely with the pipe operator, i.e., \code{|>} introduced in \proglang{R}
version~4.1.0 or \code{\%>\%} from package \pkg{magrittr} \citep{magrittr}.
Arguments \code{x}, \code{y}, \code{m}, and \code{covariates} take character,
integer, or logical vectors to select the independent variables, the dependent
variable, the hypothesized mediators, and any additional control variables,
respectively, from the data set.  Note that this interface offers various ways
to select the variables in a programmable manner.  In case of multiple
mediators, argument \code{model} allows to specify whether multiple mediators
are treated as parallel or serial mediators.

Package \pkg{robmed} provides various accessor functions to extract relevant
information from the returned objects, such as \code{coef()} and
\code{confint()} methods.  In addition, it contains the plot functions
\code{weight_plot()} and \code{ellipse_plot()} for diagnostics, as well as
\code{ci_plot()} to visualize confidence intervals and \code{density_plot()}
to plot density estimates of the indirect effect estimators.




\section[Illustrations: Using package robmed]%
{Illustrations: Using package \pkg{robmed}} \label{sec:illustrations}

We demonstrate the use of package \pkg{robmed} in three illustrative mediation
analyses using the included data set \code{BSG2014} (see
Section~\ref{sec:data}).  While the package and the data have already been
loaded in Section~\ref{sec:implementation}, we store the seed to be used for
the random number generator in an object, as it will be needed in all examples
for the purpose of replicating the results.
\begin{Schunk}
\begin{Sinput}
R> seed <- 20211117
\end{Sinput}
\end{Schunk}

The following subsections provide examples for a simple mediation
model (Section~\ref{sec:simple}), a serial multiple mediator model
(Section~\ref{sec:serial}), as well as a parallel multiple mediator model
with additional control variables (Section~\ref{sec:parallel}).

\vspace{5ex}  
\subsection{Simple mediation} \label{sec:simple}

In the first code example, we replicate parts of the empirical example of
\citet{alfons22a}.  The illustrative hypothesis to be investigated is that
task conflict mediates the relationship between team value diversity and
team commitment.  Using \pkg{robmed}'s formula interface (see
Section~\ref{sec:formula}), we specify a simple mediation model with the
dependent variable \code{TeamCommitment} on the left hand side.  On the right
hand side, we have the hypothesized mediator \code{TaskConflict}, which is
wrapped in a call to function \code{m()}, as well as the independent variable
\code{ValueDiversity}.  As we will compare the robust bootstrap test of
\citet{alfons22a} with the OLS bootstrap test \citep[e.g.,][]{preacher04,
preacher08, hayes18}, we store the formula object for later use.
\begin{Schunk}
\begin{Sinput}
R> f_simple <- TeamCommitment ~ m(TaskConflict) + ValueDiversity
\end{Sinput}
\end{Schunk}

Next, we perform the two bootstrap tests using function
\code{test_mediation()}.  As usual for functions that fit models, we supply
the model specification and the data via the \code{formula} and \code{data}
arguments.  By default, \code{test_mediation()} fits the mediation model via
regressions (argument \code{method = "regression"}) and performs a bootstrap
test for the indirect effect (argument \code{test = "boot"}) with 5000
bootstrap replications (argument \code{R = 5000}).  In that case, setting
\code{robust = TRUE} (the default) or \code{robust = "MM"} specifies the robust
bootstrap procedure of \citet{alfons22a}, while \code{robust = FALSE} yields
the nonrobust OLS bootstrap test.  Before each call to \code{test_mediation()},
we set the seed of the random number generator.
\begin{Schunk}
\begin{Sinput}
R> set.seed(seed)
R> robust_boot_simple <- test_mediation(f_simple, data = BSG2014,
+                                       robust = TRUE)
R> set.seed(seed)
R> ols_boot_simple <- test_mediation(f_simple, data = BSG2014,
+                                    robust = FALSE)
\end{Sinput}
\end{Schunk}
Other estimation methods for a bootstrap test can be specified via a
combination of arguments, as outlined in Table~\ref{tab:methods-boot}.

Function \code{test_mediation()} returns an object inheriting from class
\code{"test_mediation"}.  The corresponding \code{summary()} method shows
the relevant information on the fitted models and emphasizes the significance
tests of the total, direct, and indirect effects.  For bootstrap tests, the
information displayed by \code{summary()} by default stays within the bootstrap
framework.  For effects other than the indirect effect, asymptotic tests are
performed using the normal approximation of the bootstrap distribution.  That
is, the sample mean and the sample standard deviation of the bootstrap
replicates are used for asymptotic $z$~tests.  Furthermore, bootstrap estimates
of all effects are shown in addition to the estimates on the original data.  At
the bottom of the output, the indirect effect is summarized by the estimate on
the original data (column \code{Data}), the bootstrap estimate (i.e., the
sample mean of the bootstrap replicates; column \code{Boot}), and the lower
and upper limits of the confidence interval (columns \code{Lower} and
\code{Upper}, respectively).
\begin{Schunk}
\begin{Sinput}
R> summary(robust_boot_simple)
\end{Sinput}
\begin{Soutput}
Robust bootstrap test for indirect effect via regression

x = ValueDiversity
y = TeamCommitment
m = TaskConflict

Sample size: 89
---
Outcome variable: TaskConflict

Coefficients:
                 Data   Boot Std. Error z value Pr(>|z|)    
(Intercept)    1.1182 1.1174     0.1798   6.214 5.16e-10 ***
ValueDiversity 0.3197 0.3208     0.1071   2.996  0.00274 ** 

Robust residual standard error: 0.3033 on 87 degrees of freedom
Robust R-squared:  0.1181,	Adjusted robust R-squared:  0.108
Robust F-statistic: 9.113 on 1 and Inf DF,  p-value: 0.002539

Robustness weights:
4 observations are potential outliers with weight <= 1.3e-05:
[1] 48 58 76 79
---
Outcome variable: TeamCommitment

Coefficients:
                   Data     Boot Std. Error z value Pr(>|z|)    
(Intercept)     4.33385  4.33515    0.34415  12.597   <2e-16 ***
TaskConflict   -0.33659 -0.33672    0.17759  -1.896    0.058 .  
ValueDiversity  0.06523  0.06388    0.18593   0.344    0.731    

Robust residual standard error: 0.3899 on 86 degrees of freedom
Robust R-squared:  0.08994,	Adjusted robust R-squared:  0.06878
Robust F-statistic: 1.497 on 2 and Inf DF,  p-value: 0.2239

Robustness weights:
Observation 6 is a potential outlier with weight 0
---
Total effect of x on y:
                   Data     Boot Std. Error z value Pr(>|z|)
ValueDiversity -0.04239 -0.04293    0.18704   -0.23    0.818

Direct effect of x on y:
                  Data    Boot Std. Error z value Pr(>|z|)
ValueDiversity 0.06523 0.06388    0.18593   0.344    0.731

Indirect effect of x on y:
                Data    Boot   Lower     Upper
TaskConflict -0.1076 -0.1068 -0.2936 -0.009158
---
Level of confidence: 95 %

Number of bootstrap replicates: 5000
---
Signif. codes:  0 '***' 0.001 '**' 0.01 '*' 0.05 '.' 0.1 ' ' 1
\end{Soutput}
\begin{figure}[b!]

{\centering \includegraphics[width=0.7\textwidth]{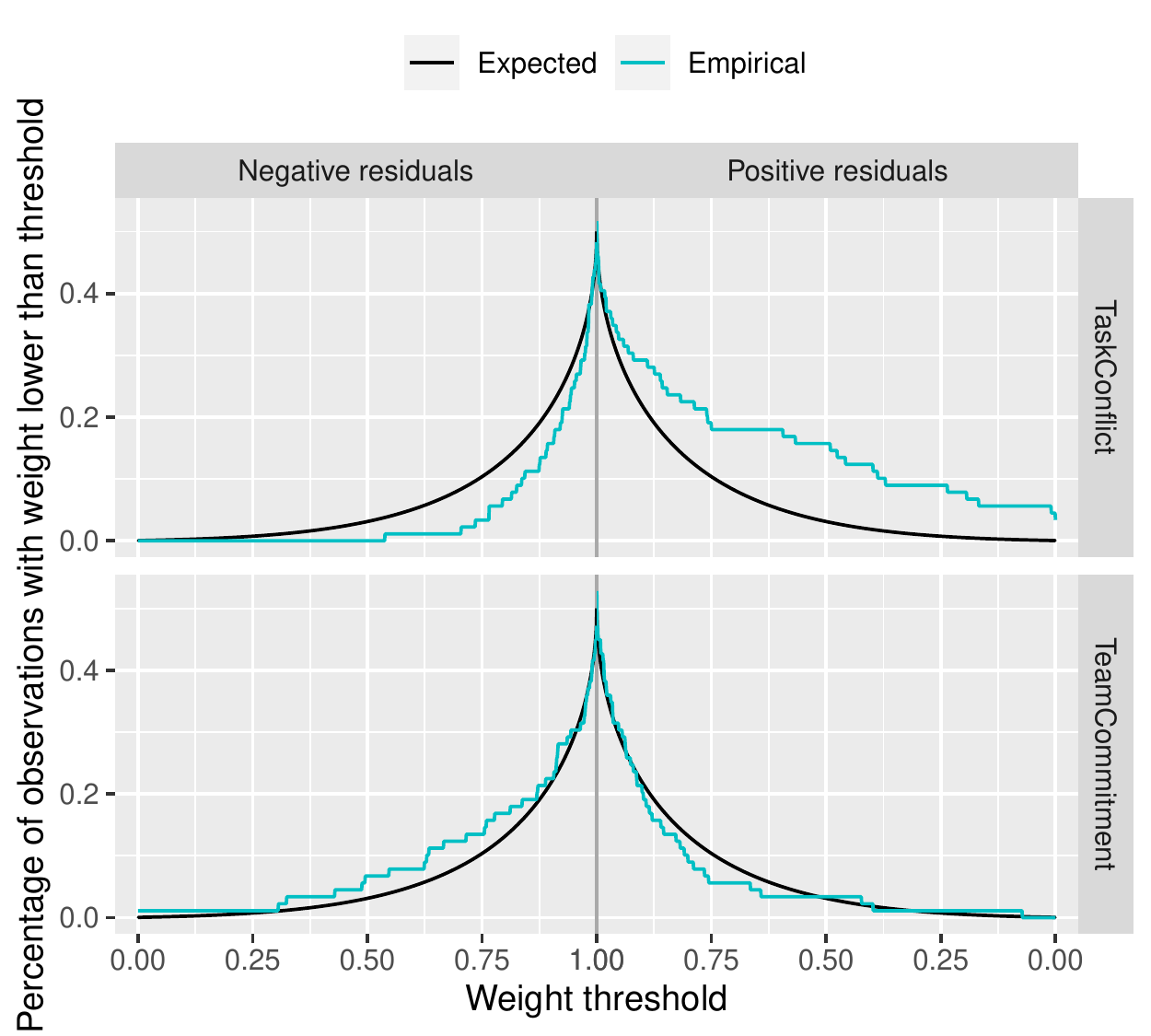} 

}

\caption[Diagnostic plot of the regression weights from the robust bootstrap procedure of \citet{alfons22a}]{Diagnostic plot of the regression weights from the robust bootstrap procedure of \citet{alfons22a}.}\label{fig:summary}
\end{figure}
\end{Schunk}
The above results are very similar to those reported in \citet{alfons22a},
but they are not identical due to a change in the random number generator in
\proglang{R} and the use of a different seed.  Specifically, the robust
bootstrap test detects a significant indirect effect, as the confidence
interval is strictly negative.  This negative indirect effect is composed of
a positive effect of value diversity on task conflict (see the output of the
first regression model), and a negative effect of task conflict on team
commitment (see the output of the second regression model).  For further
interpretation, recall that value diversity is measured as a coefficient of
variation averaged over various value dimensions, and that task conflict and
team commitment are measured as averages of items on a 5-point rating scale.
On average, an increase in value diversity by one relative standard deviation
explains an increase in task conflict by about 0.32 points, which in turn
explains a decrease in team commitment by about 0.11 points.  Furthermore, we
observe that the direct effect of value diversity on team commitment is not
significantly different from 0, meaning that value diversity affects team
commitment only via the indirect path through task conflict. In the typology
of mediations of \citet{zhao10}, we find indirect-only mediation.

Note that the output corresponding to the regression models is similar to
that of the \code{summary()} method for \code{"lmrob"} objects from package
\pkg{robustbase} \citep{robustbase}, but it is shortened as the output is
already rather long.  In particular, we emphasize that the indices of
potential outliers are displayed for each regression model.  Those potential
outliers should always be investigated further, as they may be interesting
observations that could lead to new insights when studied separately
\citep[see][for a detailed discussion on outliers in the mediation
model]{alfons22a}.

Moreover, when the summary output for the robust bootstrap procedure of
\citet{alfons22a} is printed, by default also a diagnostic plot is shown that
allows to detect deviations from normality assumptions.  Keep in mind that
this procedure uses the robust MM-estimator of regression \citep{yohai87,
salibian02}, which assigns robustness weights to all observations.  Those
weights can take any value in the interval $[0, 1]$, with lower values
indicating a higher degree of deviation.  For a varying threshold on the
horizontal axis, the diagnostic plot displays how many observations have a
weight below this threshold.  The plot is thereby split into separate panels
for negative and positive residuals.  For comparison, a reference line is
drawn for the expected percentages under normally distributed errors.

Figure~\ref{fig:summary} shows the plot for this example.  For the regression
of the hypothesized mediator (\code{TaskConflict}) on the independent variable
in the top row of the plot, it reveals much more downweighted observations with
positive residuals than expected and fewer with negative residuals.  This
indicates right skewness with a heavy upper tail.

It is possible to suppress the plot by setting \code{plot = FALSE} in
\code{summary()}.  Then function \code{weight_plot()} can be used to create
the diagnostic plot.  In this example, Figure~\ref{fig:summary} can also be
produced with the commands below.
\begin{Schunk}
\begin{Sinput}
R> weight_plot(robust_boot_simple) +
+    scale_color_manual("", values = c("black", "#00BFC4")) +
+    theme(legend.position = "top")
\end{Sinput}
\end{Schunk}

It should also be noted that the output from \code{summary()} is structured
in a similar way as the output of the widely-used macro \code{PROCESS}
\citep{hayes18}, which implements the OLS bootstrap test for conditional
process models such as the mediation model.  The intention is to facilitate
the use of package \pkg{robmed} for users of the \code{PROCESS} macro.
While \code{PROCESS} constructs a bootstrap confidence interval for the
indirect effect, it reports estimates on the original data and the usual
normal-theory $t$~tests for all other effects.  In \pkg{robmed}, the same
can be achieved by setting the argument \code{type = "data"} in the
\code{summary()} method.  The results from the regressions are then summarized
in the usual way, as shown below for the OLS bootstrap.
\begin{Schunk}
\begin{Sinput}
R> summary(ols_boot_simple, type = "data")
\end{Sinput}
\begin{Soutput}
Bootstrap test for indirect effect via regression

x = ValueDiversity
y = TeamCommitment
m = TaskConflict

Sample size: 89
---
Outcome variable: TaskConflict

Coefficients:
               Estimate Std. Error t value Pr(>|t|)    
(Intercept)      1.5007     0.2069   7.253 1.59e-10 ***
ValueDiversity   0.1552     0.1209   1.283    0.203    

Residual standard error: 0.3908 on 87 degrees of freedom
Multiple R-squared:  0.01857,	Adjusted R-squared:  0.007289
F-statistic: 1.646 on 1 and 87 DF,  p-value: 0.2029
---
Outcome variable: TeamCommitment

Coefficients:
               Estimate Std. Error t value Pr(>|t|)    
(Intercept)     4.49846    0.28806  15.616  < 2e-16 ***
TaskConflict   -0.36412    0.11783  -3.090  0.00269 ** 
ValueDiversity -0.02088    0.13418  -0.156  0.87671    

Residual standard error: 0.4296 on 86 degrees of freedom
Multiple R-squared:  0.1031,	Adjusted R-squared:  0.08227
F-statistic: 4.944 on 2 and 86 DF,  p-value: 0.009279
---
Total effect of x on y:
               Estimate Std. Error t value Pr(>|t|)
ValueDiversity -0.07738    0.13930  -0.555     0.58

Direct effect of x on y:
               Estimate Std. Error t value Pr(>|t|)
ValueDiversity -0.02088    0.13418  -0.156    0.877

Indirect effect of x on y:
                Data     Boot   Lower   Upper
TaskConflict -0.0565 -0.05838 -0.2137 0.02458
---
Level of confidence: 95 %

Number of bootstrap replicates: 5000
---
Signif. codes:  0 '***' 0.001 '**' 0.01 '*' 0.05 '.' 0.1 ' ' 1
\end{Soutput}
\end{Schunk}
Unlike the robust bootstrap test above, the OLS bootstrap does not detect a
significant indirect effect, since the confidence interval covers 0.  Due to
the influential heavy tail indicated by the diagnostic plot in
Figure~\ref{fig:summary}, the results of the robust bootstrap test can be
considered more reliable.

Methods for common generic functions to extract information from objects are
implemented in \pkg{robmed}, such as a \code{coef()} method to extract the
relevant effects of the mediation model, and \code{confint()} to extract
confidence intervals of those effects.
\begin{Schunk}
\begin{Sinput}
R> coef(robust_boot_simple)
\end{Sinput}
\begin{Soutput}
          a           b       Total      Direct    Indirect 
 0.32077184 -0.33672132 -0.04292728  0.06388476 -0.10681204 
\end{Soutput}
\begin{Sinput}
R> confint(robust_boot_simple)
\end{Sinput}
\begin{Soutput}
              2.5 %       97.5 %
a         0.1109227  0.530620997
b        -0.6847919  0.011349287
Total    -0.4095137  0.323659113
Direct   -0.3005305  0.428300051
Indirect -0.2936367 -0.009158447
\end{Soutput}
\end{Schunk}
While the confidence intervals in this example do not add much in terms
of interpretation over the output of \code{summary()}, the latter reports
significance tests instead of confidence intervals for the effects other
than the indirect effect.  For researchers who prefer to report confidence
intervals, the \code{confint()} method allows to easily extract this
information.

For objects corresponding to bootstrap tests (class
\code{"boot_test_mediation"}), argument \code{type} of the \code{coef()} method
allows to specify whether to extract the bootstrap estimates (\code{"boot"},
the default) or the estimates on the original data (\code{"data"}).  Similarly,
argument \code{type} of the \code{confint()} method allows to specify whether
the confidence intervals for the effects other than the indirect effect should
be bootstrap confidence intervals (\code{"boot"}, the default) or normal theory
intervals based on the original data (\code{"data"}).
\begin{Schunk}
\begin{Sinput}
R> coef(ols_boot_simple, type = "data")
\end{Sinput}
\begin{Soutput}
          a           b       Total      Direct    Indirect 
 0.15517748 -0.36412398 -0.07738318 -0.02087934 -0.05650384 
\end{Soutput}
\begin{Sinput}
R> confint(ols_boot_simple, type = "data")
\end{Sinput}
\begin{Soutput}
               2.5 %     97.5 %
a        -0.08521602  0.3955710
b        -0.59836363 -0.1298843
Total    -0.35426561  0.1994992
Direct   -0.28761575  0.2458571
Indirect -0.21371944  0.0245833
\end{Soutput}
\end{Schunk}
In addition, argument \code{parm} allows to specify which coefficients or
confidence intervals to extract.
\begin{Schunk}
\begin{Sinput}
R> coef(robust_boot_simple, parm = "Indirect")
\end{Sinput}
\begin{Soutput}
 Indirect 
-0.106812 
\end{Soutput}
\begin{Sinput}
R> confint(robust_boot_simple, parm = "Indirect")
\end{Sinput}
\begin{Soutput}
              2.5 %       97.5 %
Indirect -0.2936367 -0.009158447
\end{Soutput}
\end{Schunk}
While the bootstrap tests implemented in \code{test_mediation()} construct a
confidence interval for the indirect effect based on a pre-specified confidence
level $1-\alpha$, function \code{p_value()} allows to analyze the bootstrap
distribution and extract the smallest significance level $\alpha$ for which the
$(1-\alpha)\cdot 100\%$ confidence interval of the indirect effect does not
contain 0.
\begin{Schunk}
\begin{Sinput}
R> p_value(robust_boot_simple, parm = "Indirect")
\end{Sinput}
\begin{Soutput}
Indirect 
  0.0281 
\end{Soutput}
\begin{Sinput}
R> p_value(ols_boot_simple, parm = "Indirect")
\end{Sinput}
\begin{Soutput}
Indirect 
  0.1483 
\end{Soutput}
\end{Schunk}
Here, the $p$~value for the robust bootstrap test shows strong evidence against
the null hypothesis of no mediation, whereas the OLS bootstrap fails to do so.

Several plots are implemented to visualize the results of mediation analysis.
They can be applied to each object individually, but it is also possible to
combine multiple objects from mediation analysis into a list in order to
compare different methods graphically.  If names are given to the list
elements, those names will be used by the plots to identify the different
methods.  Note that we use the name \code{"ROBMED"} here to refer to the robust
bootstrap test.
\begin{Schunk}
\begin{Sinput}
R> boot_list <- list("OLS bootstrap" = ols_boot_simple,
+                    "ROBMED" = robust_boot_simple)
\end{Sinput}
\end{Schunk}

Function \code{density_plot()} plots the density estimates of the indirect
effect.  It also adds vertical lines for the point estimates and illustrates
the confidence intervals by shaded areas.  The plot resulting from the
following command is displayed in Figure~\ref{fig:density}.
\begin{Schunk}
\begin{Sinput}
R> density_plot(boot_list)
\end{Sinput}
\begin{figure}[t!]

{\centering \includegraphics[width=0.7\textwidth]{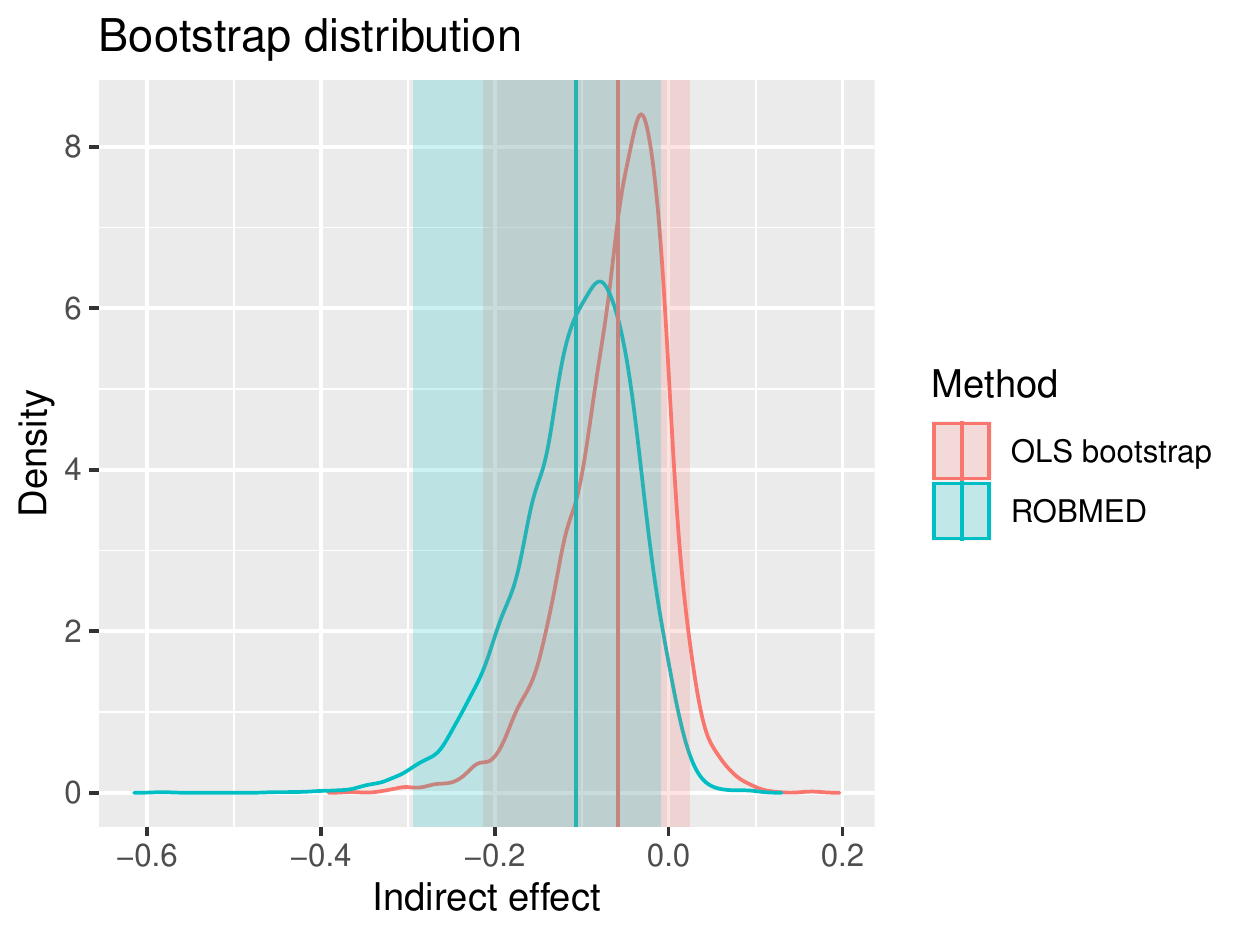} 

}

\caption[Density plot of the bootstrap distributions of the indirect effect, obtained via the OLS bootstrap and the robust bootstrap procedure of \citet{alfons22a}]{Density plot of the bootstrap distributions of the indirect effect, obtained via the OLS bootstrap and the robust bootstrap procedure of \citet{alfons22a}.  The vertical lines indicate the the respective point estimates of the indirect effect and the shaded areas represent the confidence intervals.}\label{fig:density}
\end{figure}
\end{Schunk}

In order to aid with interpretation of the results from mediation analysis,
function \code{ci_plot()} allows to visualize the point estimates and
confidence intervals of selected effects. The direct effect and the indirect
effect are plotted by default, as the typology of mediations of \citet{zhao10}
is based on those two effects.  Nevertheless, argument \code{parm} can be used
to specify which effects to plot.  Figure~\ref{fig:ci} contains the plot
created by the command below.
\begin{Schunk}
\begin{Sinput}
R> ci_plot(boot_list, parm = c("a", "b", "Direct", "Indirect"))
\end{Sinput}
\begin{figure}[t!]

{\centering \includegraphics[width=0.85\textwidth]{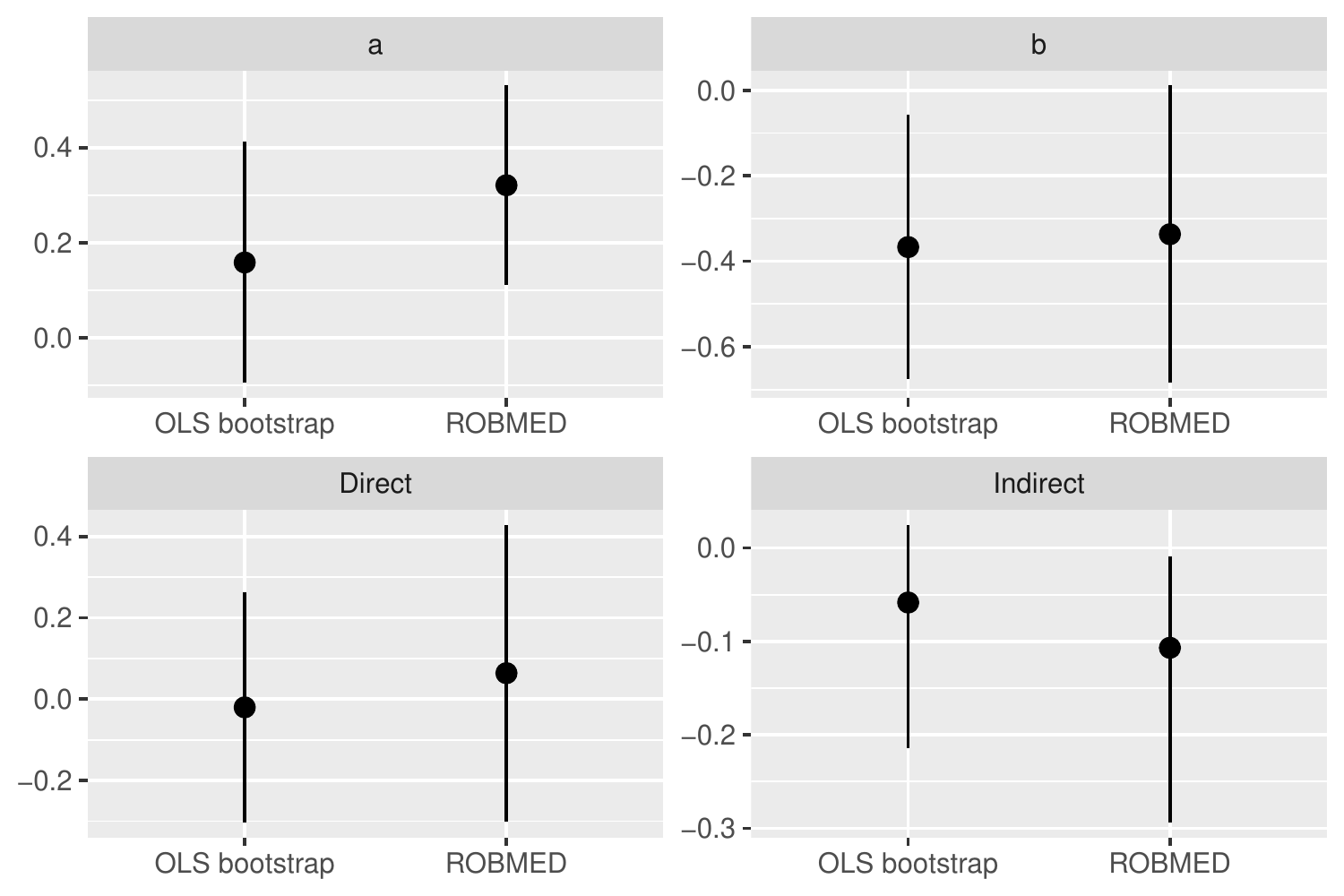} 

}

\caption[Point estimates and 95\% confidence intervals for selected effects in the mediation model, estimated via the OLS bootstrap and the robust bootstrap procedure of \citet{alfons22a}]{Point estimates and 95\% confidence intervals for selected effects in the mediation model, estimated via the OLS bootstrap and the robust bootstrap procedure of \citet{alfons22a}.}\label{fig:ci}
\end{figure}
\end{Schunk}

Finally, function \code{ellipse_plot()} produces a bivariate scatterplot
together with a tolerance ellipse that illustrates how well the regression
results represent the data, exploiting the relationship between regression
coefficients and the covariance matrix.  For the robust bootstrap procedure of
\citet{alfons22a}, the robustness weights from the robust regression estimator
can be used to compute a weighted sample covariance matrix, from which the
tolerance ellipse is computed.  It is important to note that such a weighted
covariance matrix is not Fisher consistent (that is, the functional form of the
estimator applied to the model distribution does not equal the true covariance
matrix), as observations are also expected to be downweighted when all
observations follow the model. However, it is straightforward to obtain a
correction for a Fisher consistent covariance matrix (see
Appendix~\ref{app:ellipse}).

For instance, we can produce such a plot with the independent variable on the
horizontal axis and the hypothesized mediator on the vertical axis.  In that
case the plot represents the results from the regression of the hypothesized
mediator on the independent variable.  As the independent variable is the only
explanatory variable in this regression model, the plot also adds lines
representing the respective regression coefficients.
\begin{Schunk}
\begin{Sinput}
R> ellipse_plot(boot_list, horizontal = "ValueDiversity",
+               vertical = "TaskConflict")
\end{Sinput}
\begin{figure}[b!]

{\centering \includegraphics[width=0.7\textwidth]{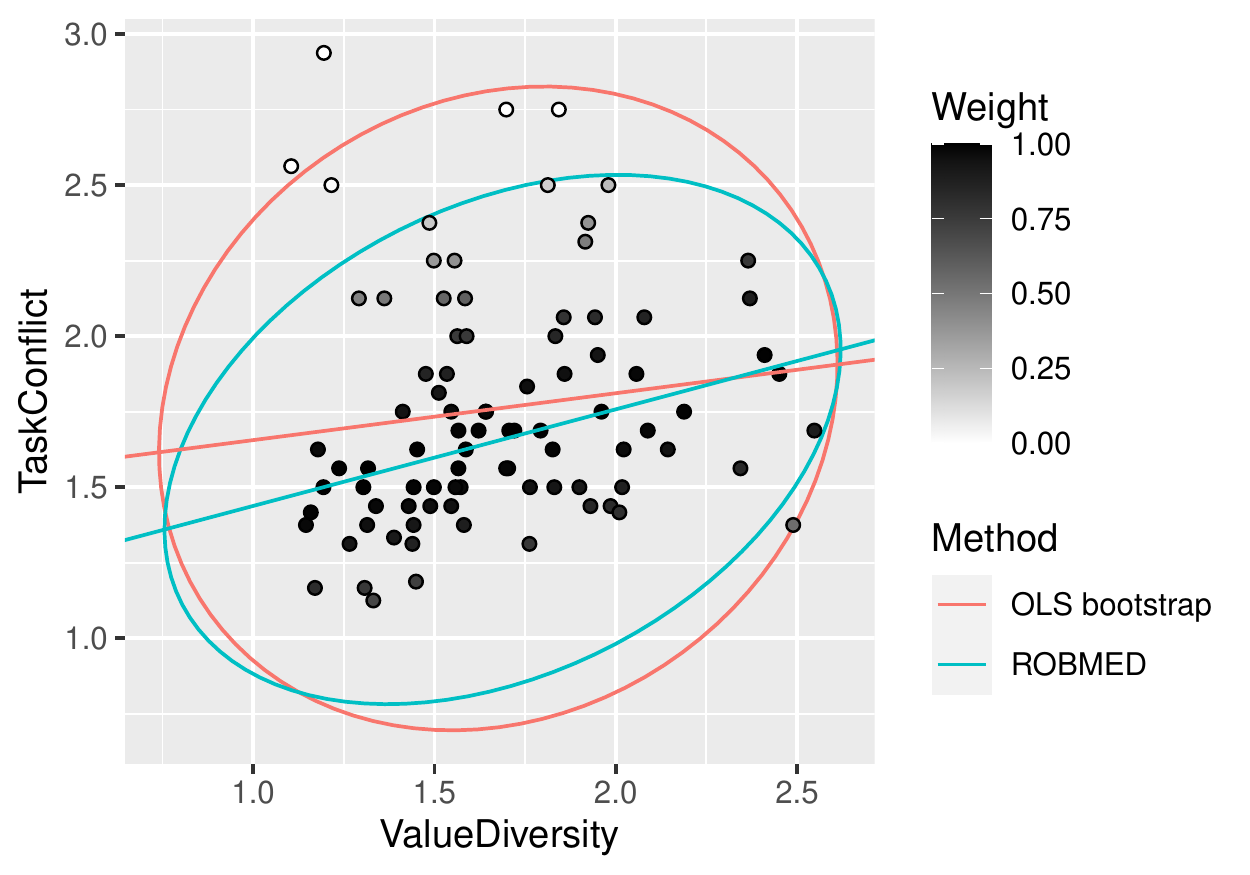} 

}

\caption[Diagnostic plot with tolerance ellipses for the OLS bootstrap and the robust bootstrap procedure of \citet{alfons22a}]{Diagnostic plot with tolerance ellipses for the OLS bootstrap and the robust bootstrap procedure of \citet{alfons22a}.}\label{fig:ellipse}
\end{figure}
\end{Schunk}
The resulting plot is shown in Figure~\ref{fig:ellipse}.  Since we are
comparing a nonrobust method with a robust method that assigns a robustness
weight to each observation, by default those robustness weights are
visualized by plotting the points on a grayscale.  Clearly, the tolerance
ellipse corresponding to the robust method fits the main bulk of the data
better, as the tolerance ellipse corresponding to the nonrobust method
contains more empty space.  This plot further suggests that the detected
potential outliers (see also the printed output of \code{summary()} above)
are a result of the heavy upper tail in the hypothesized mediator
(\code{TaskConflict}).

All plot functions in \pkg{robmed} allow customization via the underlying
package \pkg{ggplot2} \citep{wickham16}.  Arguments can be passed down to
\code{geom_xxx()} functions, and additional elements can be added to the plot
as usual with the \code{+} operator.  We refer to the \proglang{R} help
files of the plots for some examples.  Nevertheless, there are limits to the
customization, in particular for the plots that contain various elements such
as the diagnostic plot with the tolerance ellipse.

For further customization, \pkg{robmed} provides the workhorse functions
\code{setup_weight_plot()},  \code{setup_ci_plot()},
\code{setup_density_plot()}, and \code{setup_ellipse_plot()}.  They
do not produce the plot, but they extract the relevant information to be
displayed.  They are useful for users who want to create similar plots, but
who want more control over what information to display or how to display that
information.  With the commands below, we manually produce the same plot as
before, but only plot the tolerance ellipses and the data without the
regression lines.  Figure~\ref{fig:ellipse-custom} displays the resulting plot.
\begin{Schunk}
\begin{Sinput}
R> setup <- setup_ellipse_plot(boot_list, horizontal = "ValueDiversity",
+                              vertical = "TaskConflict")
R> ggplot() +
+    geom_path(aes(x = x, y = y, color = Method), data = setup$ellipse) +
+    geom_point(aes(x = x, y = y, fill = Weight), data = setup$data,
+               shape = 21) +
+    scale_fill_gradient(limits = 0:1, low = "white", high = "black") +
+    labs(x = setup$horizontal, y = setup$vertical)
\end{Sinput}
\begin{figure}[t]

{\centering \includegraphics[width=0.7\textwidth]{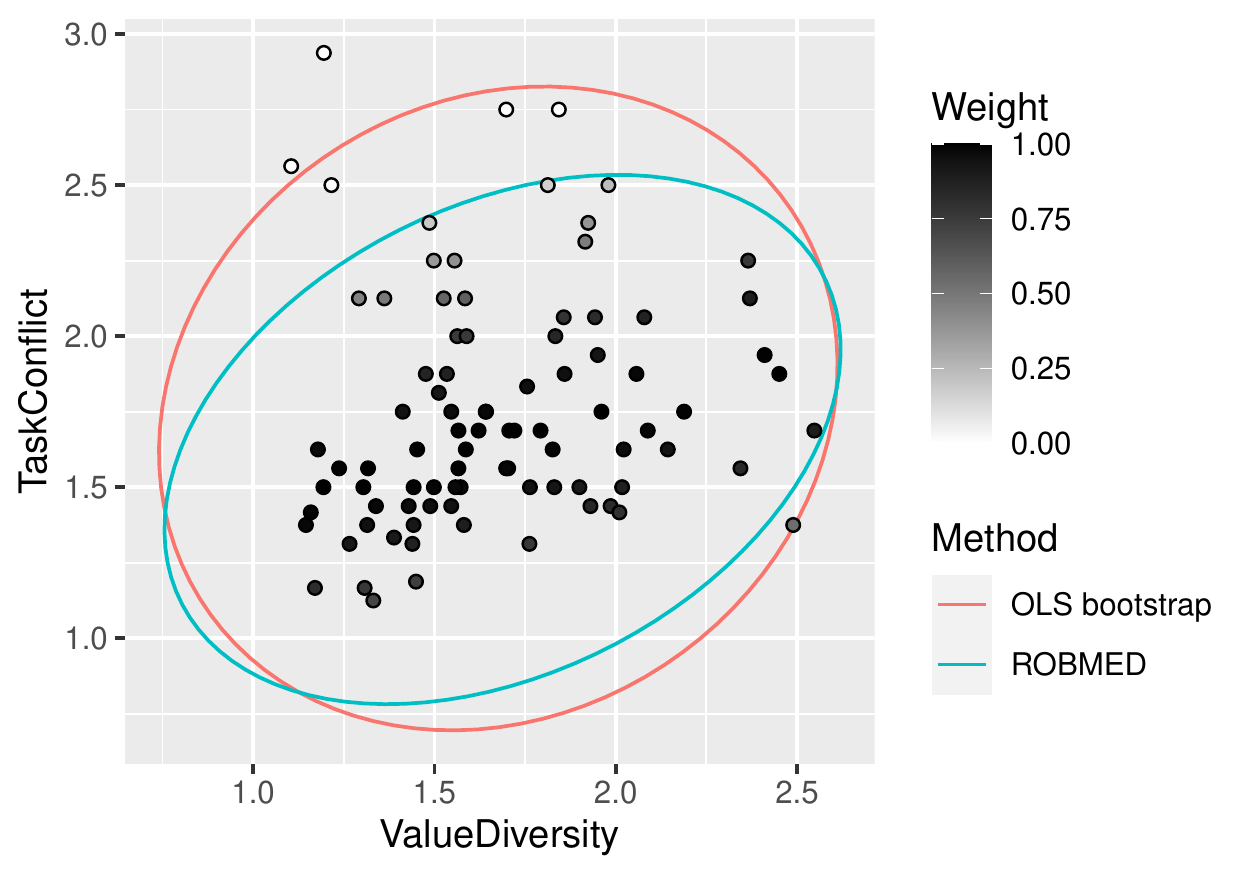} 

}

\caption[Customized diagnostic plot with tolerance ellipses but without regression lines for the OLS bootstrap and the robust bootstrap procedure of \citet{alfons22a}]{Customized diagnostic plot with tolerance ellipses but without regression lines for the OLS bootstrap and the robust bootstrap procedure of \citet{alfons22a}.}\label{fig:ellipse-custom}
\end{figure}
\end{Schunk}

\subsection{Serial multiple mediators}
\label{sec:serial}

The second example extends the simple mediation model from the previous
section to a serial multiple mediator model.  We investigate the following
illustrative hypothesis: value diversity negatively impacts team commitment
through increased task conflict, and in turn task conflict negatively affects
team performance through decreased team commitment.  Here the dependent
variable is an objective assessment of team performance, measured by the
team's score on the simulation game.

The corresponding formula is stored in an object for later use and contains the
dependent variable \code{TeamScore} on the left hand side.  On the right hand
side, the hypothesized serial mediators \code{TaskConflict} and
\code{TeamCommitment} are wrapped in a call to \code{serial_m()}, separated by
the usual \code{+} operator from the independent variable \code{ValueDiversity}.
\begin{Schunk}
\begin{Sinput}
R> f_serial <- TeamScore ~ serial_m(TaskConflict, TeamCommitment) +
+    ValueDiversity
\end{Sinput}
\end{Schunk}
Function \code{test_mediation()} is then called in the same way as in the
previous section to compare the robust bootstrap procedure of \citet{alfons22a}
with the nonrobust OLS bootstrap.
\begin{Schunk}
\begin{Sinput}
R> set.seed(seed)
R> robust_boot_serial <- test_mediation(f_serial, data = BSG2014,
+                                       robust = TRUE)
R> set.seed(seed)
R> ols_boot_serial <- test_mediation(f_serial, data = BSG2014,
+                                    robust = FALSE)
\end{Sinput}
\end{Schunk}

The output of \code{summary()} looks very similar to that of the previous
example, except that the last part shows results for multiple indirect
effects.  In order to save space, we only print the objects returned by
\code{test_mediation()}, which shows the results for the bootstrap confidence
intervals of the indirect effects.  For completeness, the full \code{summary()}
output of the robust bootstrap test can be found in Appendix~\ref{app:output}.
\begin{Schunk}
\begin{Sinput}
R> robust_boot_serial
\end{Sinput}
\begin{Soutput}
Robust bootstrap tests for indirect effects via regression

Indirect effects of x on y:
             Data     Boot  Lower    Upper
Total     -0.2870 -0.46185 -6.859  2.87613
Indirect1  0.1036  0.07264 -1.668  2.24594
Indirect2  0.6010  0.45184 -3.130  5.13150
Indirect3 -0.9916 -0.98633 -3.909 -0.06841

Indirect effect paths:
 Label     Path                                                           
 Indirect1 ValueDiversity -> TaskConflict   -> TeamScore                  
 Indirect2 ValueDiversity -> TeamCommitment -> TeamScore                  
 Indirect3 ValueDiversity -> TaskConflict   -> TeamCommitment -> TeamScore
---
Level of confidence: 95 %

Number of bootstrap replicates: 5000
\end{Soutput}
\begin{Sinput}
R> ols_boot_serial
\end{Sinput}
\begin{Soutput}
Bootstrap tests for indirect effects via regression

Indirect effects of x on y:
               Data     Boot  Lower   Upper
Total     -0.427908 -0.52270 -3.530 1.60164
Indirect1  0.009045 -0.11381 -1.179 1.24710
Indirect2 -0.117898 -0.09866 -2.462 1.61605
Indirect3 -0.319055 -0.31024 -1.518 0.08132

Indirect effect paths:
 Label     Path                                                           
 Indirect1 ValueDiversity -> TaskConflict   -> TeamScore                  
 Indirect2 ValueDiversity -> TeamCommitment -> TeamScore                  
 Indirect3 ValueDiversity -> TaskConflict   -> TeamCommitment -> TeamScore
---
Level of confidence: 95 %

Number of bootstrap replicates: 5000
\end{Soutput}
\end{Schunk}
Note that the row labeled \code{Total} in the above output contains the results
for the sum of the three individual indirect effects.  We observe that the
robust bootstrap test detects mediation in the indirect path that goes first
through task conflict and subsequently through team commitment (effect
\code{Indirect3}), whereas none of the indirect effects are significant in the
OLS bootstrap.

We can use function \code{weight_plot()} to produce a diagnostic plot to
investigate deviations from normality assumptions that could explain the
differences in results for the two methods.  The plot created with the
commands below is shown in Figure~\ref{fig:weight}.
\begin{Schunk}
\begin{Sinput}
R> weight_plot(robust_boot_serial) +
+    scale_color_manual("", values = c("black", "#00BFC4")) +
+    theme(legend.position = "top")
\end{Sinput}
\begin{figure}[t]

{\centering \includegraphics[width=0.7\textwidth]{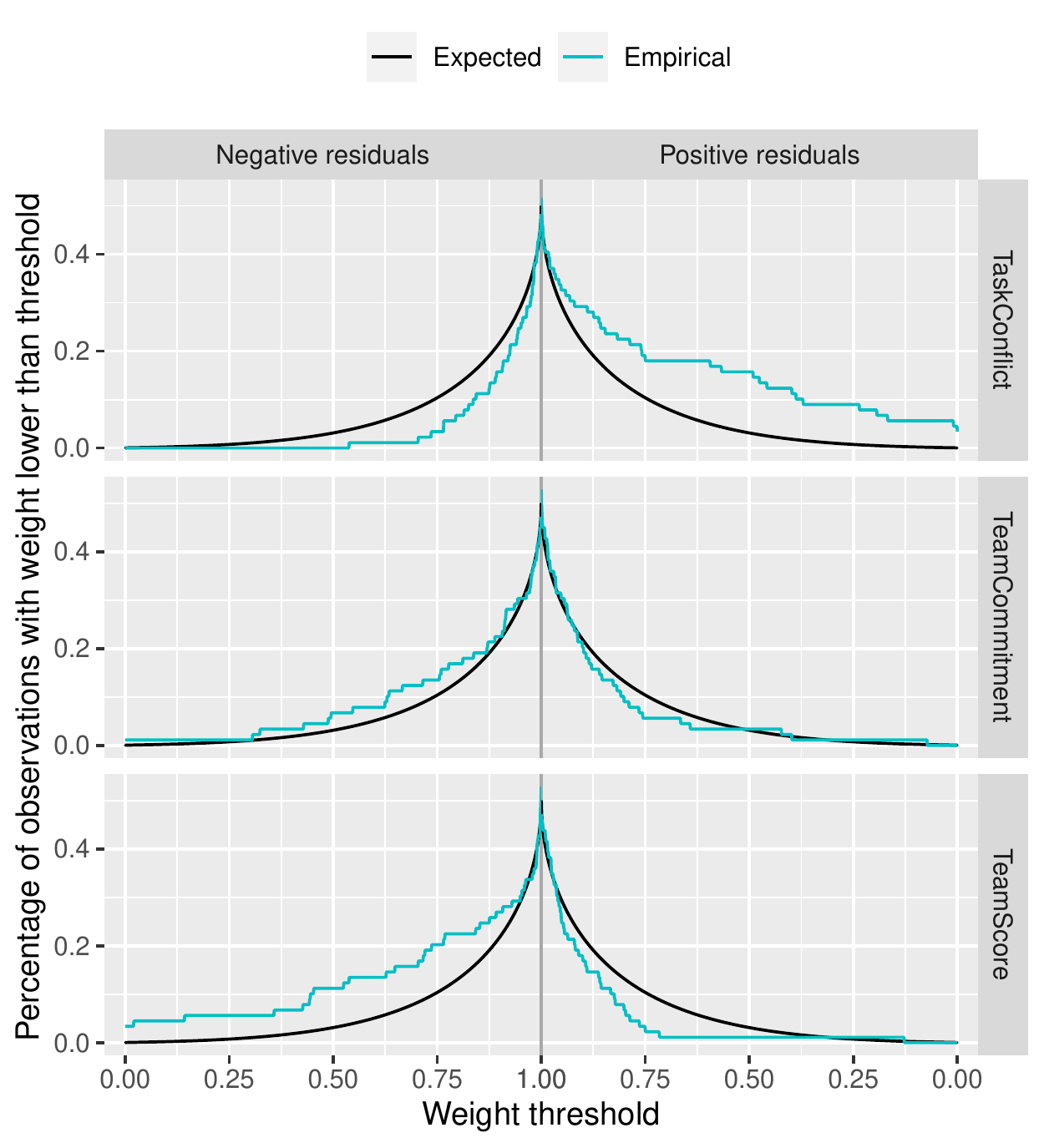} 

}

\caption[Diagnostic plot of the regression weights from the robust bootstrap procedure of \citet{alfons22a} in the example for a serial multiple mediator model]{Diagnostic plot of the regression weights from the robust bootstrap procedure of \citet{alfons22a} in the example for a serial multiple mediator model.}\label{fig:weight}
\end{figure}
\end{Schunk}
As in the previous example, we see that there are strong indications of a heavy
upper tail in the regression of task conflict on value diversity (top row of
Figure~\ref{fig:weight}) and a heavy lower tail in the regression for objective
team performance (bottom row of Figure~\ref{fig:weight}), which makes the
results of the OLS bootstrap unreliable.

\subsection{Parallel multiple mediators and control variables}
\label{sec:parallel}

In the final example, we investigate the illustrative hypothesis that
procedural justice and interactional justice are parallel mediators for the
relationship between shared leadership and the team's perception of its
performance, controlled for diversity in age and gender within the team.
Unlike in the previous section, the dependent variable is a subjective measure
of team performance, as evaluated by the team members themselves.

In \pkg{robmed}'s formula interface, parallel multiple mediators can be
specified by wrapping multiple variables in a call to function
\code{parallel_m()}, and control variables can be specified in a similar
manner with function \code{covariates()}.
\begin{Schunk}
\begin{Sinput}
R> f_parallel <-
+    TeamPerformance ~ parallel_m(ProceduralJustice, InteractionalJustice) +
+    SharedLeadership + covariates(AgeDiversity, GenderDiversity)
\end{Sinput}
\end{Schunk}

Function \code{test_mediation()} is then called in the usual way to compare the
robust bootstrap test of \citet{alfons22a} and the nonrobust OLS bootstrap.
\begin{Schunk}
\begin{Sinput}
R> set.seed(seed)
R> robust_boot_parallel <- test_mediation(f_parallel, data = BSG2014,
+                                         robust = TRUE)
R> set.seed(seed)
R> ols_boot_parallel <- test_mediation(f_parallel, data = BSG2014,
+                                      robust = FALSE)
\end{Sinput}
\end{Schunk}
To save space, we again do not show the entire \code{summary()} of the
resulting objects, but only the results for the indirect effect by printing
the objects.  Appendix~\ref{app:output} contains the full \code{summary()}
output of the robust bootstrap test, including the diagnostic plot of the
regression weights.
\begin{Schunk}
\begin{Sinput}
R> robust_boot_parallel
\end{Sinput}
\begin{Soutput}
Robust bootstrap tests for indirect effects via regression

Indirect effects of x on y:
                        Data    Boot     Lower  Upper
Total                0.12089 0.12546 4.398e-02 0.2164
ProceduralJustice    0.03984 0.04568 6.206e-05 0.1184
InteractionalJustice 0.08105 0.07978 3.022e-02 0.1538
---
Level of confidence: 95 %

Number of bootstrap replicates: 5000
\end{Soutput}
\begin{Sinput}
R> ols_boot_parallel
\end{Sinput}
\begin{Soutput}
Bootstrap tests for indirect effects via regression

Indirect effects of x on y:
                        Data    Boot    Lower   Upper
Total                0.07452 0.07696 0.015240 0.14838
ProceduralJustice    0.04093 0.04631 0.004064 0.10983
InteractionalJustice 0.03359 0.03064 0.004396 0.08471
---
Level of confidence: 95 %

Number of bootstrap replicates: 5000
\end{Soutput}
\end{Schunk}
As for the serial multiple mediator model, the row \code{Total} in the output
above contains the results for the sum of the two individual indirect effects
through the hypothesized mediators \code{ProceduralJustice} and
\code{InteractionalJustice}.  We observe that both individual indirect effects
are significant at the 5\% level for both the robust procedure of
\citet{alfons22a} and the OLS bootstrap.  However, the lower confidence bound
for the indirect effect of procedural justice in the robust bootstrap is so
small that different seeds of the random number generator often yield a
(similarly small) negative lower bound.  Overall, we may conclude that we find
at least weak support for mediation via procedural justice, and we do find
support for mediation via interactional justice.

The output of \code{summary()} in Appendix~\ref{app:output}, indicates that
the residual distributions are fairly normal (see the diagnostic plot in
Figure~\ref{fig:summary-parallel}), but that there is a small number of
potential outliers that should be investigated further.  As the regression of
the dependent variable (\code{TeamPerformance}) on the remaining variables
shows more deviations from normality than the other regressions (bottom row of
Figure~\ref{fig:summary-parallel}), we use \code{ellipse_plot()} to create the
diagnostic plot with a tolerance ellipse that is related to the regression
results.  If several explanatory variables are included, it is often more
insightful to plot the partial residuals on the vertical axis, i.e., to
subtract from the response the linear predictor without the variable that is
displayed on the horizontal axis.  This has the additional advantage that the
regression coefficient can be visualized by a line, which is otherwise not
possible in the case of multiple explanatory variables.  With function
\code{ellipse_plot()}, plotting the partial residuals can easily be achieved
by setting the argument \code{partial = TRUE}.  With the command below, we plot
the partial residuals of team performance against the independent variable
shared leadership.  Figure~\ref{fig:ellipse-partial} contains the resulting
plot, which clearly visualizes the noisy data points.
\begin{Schunk}
\begin{Sinput}
R> ellipse_plot(robust_boot_parallel, horizontal = "SharedLeadership",
+               vertical = "TeamPerformance", partial = TRUE)
\end{Sinput}
\begin{figure}[t]

{\centering \includegraphics[width=0.7\textwidth]{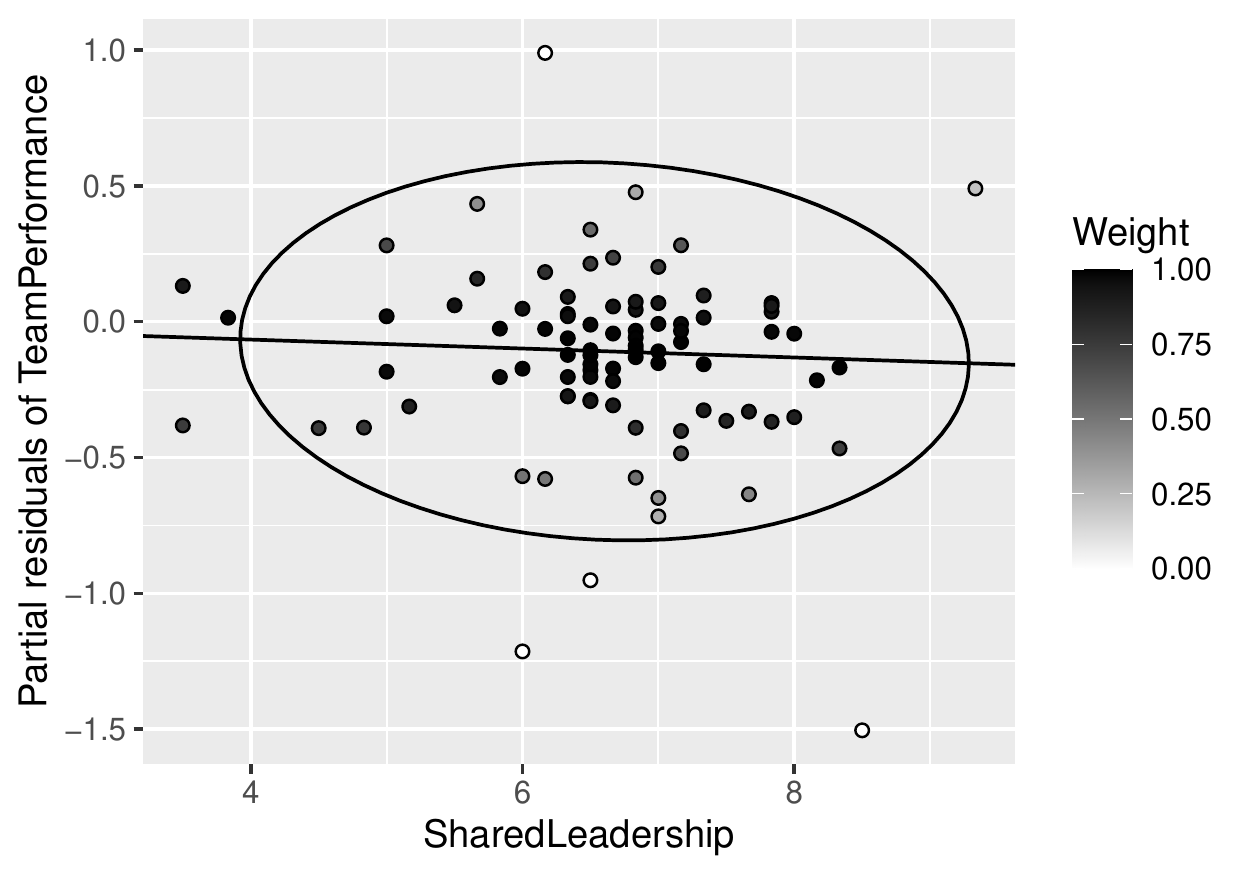} 

}

\caption[Diagnostic plot with a tolerance ellipse for partial residuals in a multiple parallel mediator model]{Diagnostic plot with a tolerance ellipse for partial residuals in a multiple parallel mediator model.}\label{fig:ellipse-partial}
\end{figure}
\end{Schunk}

In multiple mediator models, it can be of interest if the indirect effects are
different from one another, or if they differ in magnitude
\citep[p.163--166]{hayes18}.  Function \code{test_mediation()} allows to make
pairwise comparisons of indirect effects via the argument \code{contrast}.  By
setting this argument to \code{"estimates"}, the pairwise differences of the
estimates of the indirect effect are computed, whereas setting this argument to
\code{"absolute"} yields the computation of pairwise differences in absolute
values.  The output of \code{test_mediation()} then includes the estimates and
confidence intervals for those contrasts, and also displays information on the
definition of the contrasts.
\begin{Schunk}
\begin{Sinput}
R> set.seed(seed)
R> test_mediation(f_parallel, data = BSG2014, contrast = "absolute")
\end{Sinput}
\begin{Soutput}
Robust bootstrap tests for indirect effects via regression

Indirect effects of x on y:
                         Data     Boot      Lower   Upper
Total                 0.12089  0.12546  4.398e-02 0.21640
ProceduralJustice     0.03984  0.04568  6.206e-05 0.11841
InteractionalJustice  0.08105  0.07978  3.022e-02 0.15384
Contrast             -0.04121 -0.03393 -1.164e-01 0.05817

Indirect effect contrast definition:
 Label    Definition                                  
 Contrast |ProceduralJustice| - |InteractionalJustice|
---
Level of confidence: 95 %

Number of bootstrap replicates: 5000
\end{Soutput}
\end{Schunk}

However, it is not actually necessary to run the entire bootstrap procedure
again if a bootstrap test had already been performed without computing those
contrasts.  Function \code{retest()} allows to reanalyze the bootstrap
estimates with different parameter settings, which saves computation time in
such a case.
\begin{Schunk}
\begin{Sinput}
R> retest(robust_boot_parallel, contrast = "absolute")
\end{Sinput}
\begin{Soutput}
Robust bootstrap tests for indirect effects via regression

Indirect effects of x on y:
                         Data     Boot      Lower   Upper
Total                 0.12089  0.12546  4.398e-02 0.21640
ProceduralJustice     0.03984  0.04568  6.206e-05 0.11841
InteractionalJustice  0.08105  0.07978  3.022e-02 0.15384
Contrast             -0.04121 -0.03393 -1.164e-01 0.05817

Indirect effect contrast definition:
 Label    Definition                                  
 Contrast |ProceduralJustice| - |InteractionalJustice|
---
Level of confidence: 95 %

Number of bootstrap replicates: 5000
\end{Soutput}
\end{Schunk}
We emphasize that function \code{retest()} is available for computational
convenience in case the analysis was by mistake conducted with the wrong
parameter settings.  It must not be abused for $p$~hacking.


\section{Summary and discussion} \label{sec:summary}

The \proglang{R} package \pkg{robmed} provides easy-to-use functionality for
robust mediation analysis.  It implements the robust bootstrap procedure of
\citet{alfons22a}, which yields reliable results under outliers and other
deviations from normality assumptions, as well as diagnostic plots that allow
to detect and further investigate such deviations.  In addition, package
\pkg{robmed} provides functionality for various other procedures for mediation
analysis, as well as plots that allow to visualize and compare results from
different methods.  All implemented methods thereby share the same function
interface and a clear class structure of the results.  In particular,
\pkg{robmed} introduces a new formula interface that allows to specify
various types of mediation models with a single formula.

At present, there are some limitations of package \pkg{robmed}.  First of
all, the current version requires a numeric dependent variable and numeric
mediators, although the independent variable and additional control variables
may be binary or categorical.  We aim to extend the fast-and-robust bootstrap
methodology for robust estimators of logistic regression in order to add
support for mediation analysis with binary dependent variables.  Second,
adding support for additional mediation models, such as moderated mediation
and mediated moderation models \citep[e.g.,][]{muller05} is planned for future
versions.  Third, the diagnostic plot of the regression weights would be even
more useful if it included confidence bands, which could perhaps be constructed
from the bootstrap procedure.  Finally, a graphical user interface (GUI) could
be beneficial for less proficient \proglang{R} users.  Developing such a GUI,
for instance as a web application based on package \pkg{shiny} \citep{shiny},
is future work.  In the meantime, we provide the \proglang{R} extension bundle
\pkg{ROBMED} \citep{ROBMED-RSPSS} for \proglang{SPSS} \citep{SPSS}, which links
to the \proglang{R} package \pkg{robmed} and allows to use its main
functionality through a GUI from within \proglang{SPSS}.  Interested
readers can obtain the extension bundle from
\url{https://github.com/aalfons/ROBMED-RSPSS}.


\section*{Computational details}


The results in this paper were obtained using \proglang{R}
version~4.2.1
\citep{R} with package \pkg{robmed} version~1.0.0
\citep{robmed} and its dependencies \pkg{boot}
version~1.3.28 \citep{boot},
\pkg{ggplot2} version~3.3.6 \citep{wickham16},
and \pkg{robustbase} version~0.95.0
\citep{robustbase}.  Package \pkg{robmed} also builds upon packages
\pkg{quantreg} \citep{quantreg} and \pkg{sn} \citep{sn} for functionality not
shown in this paper, as well as package \pkg{testthat} \citep{wickham11} for
unit tests.  \proglang{R} itself and all mentioned packages are available from
the Comprehensive \proglang{R} Archive Network (CRAN) at
\url{https://CRAN.R-project.org/}.  The latest development version of package
\pkg{robmed} can be obtained from \url{https://github.com/aalfons/robmed}.

\section*{Acknowledgments}


Andreas Alfons is supported by a grant of the Dutch Research Council (NWO),
research program Vidi, project number \mbox{VI.Vidi.195.141}.  N\"{u}fer
Y.\ Ate\c{s} is supported by the Science Academy Young Scientists Award
Program (BAGEP) of the Science Academy Society of Turkey.


\vspace{5ex}  
\bibliography{robmed}

\begin{thebibliography}{63}
\newcommand{\enquote}[1]{``#1''}
\providecommand{\natexlab}[1]{#1}
\providecommand{\url}[1]{\texttt{#1}}
\providecommand{\urlprefix}{URL }
\expandafter\ifx\csname urlstyle\endcsname\relax
  \providecommand{\doi}[1]{doi:\discretionary{}{}{}#1}\else
  \providecommand{\doi}{doi:\discretionary{}{}{}\begingroup
  \urlstyle{rm}\Url}\fi
\providecommand{\eprint}[2][]{\url{#2}}

\bibitem[{Alfons(2022{\natexlab{a}})}]{ROBMED-RSPSS}
Alfons A (2022{\natexlab{a}}).
\newblock \emph{\pkg{ROBMED}: \proglang{SPSS} \proglang{R} Extension Bundle for
  Robust Mediation Analysis}.
\newblock \doi{10.25397/eur.19354268.v1}.
\newblock \urlprefix\url{https://github.com/aalfons/ROBMED-RSPSS}.

\bibitem[{Alfons(2022{\natexlab{b}})}]{robmed}
Alfons A (2022{\natexlab{b}}).
\newblock \emph{\pkg{robmed}: (Robust) Mediation Analysis}.
\newblock \proglang{R}~package version~1.0.0,
  \urlprefix\url{https://CRAN.R-project.org/package=robmed}.

\bibitem[{Alfons \emph{et~al.}(2022)Alfons, Ate\c{s}, and Groenen}]{alfons22a}
Alfons A, Ate\c{s} NY, Groenen PJF (2022).
\newblock \enquote{A Robust Bootstrap Test for Mediation Analysis.}
\newblock \emph{Organizational Research Methods}, \textbf{25}(3), 591--617.
\newblock \doi{10.1177/1094428121999096}.

\bibitem[{Asparouhov and Muth\'{e}n(2016)}]{asparouhov16}
Asparouhov T, Muth\'{e}n B (2016).
\newblock \enquote{Structural Equation Models and Mixture Models with
  Continuous Nonnormal Skewed Distributions.}
\newblock \emph{Structural Equation Modeling: A Multidisciplinary Journal},
  \textbf{23}(1), 1--19.
\newblock \doi{10.1080/10705511.2014.947375}.

\bibitem[{Azzalini(2022)}]{sn}
Azzalini A (2022).
\newblock \emph{\pkg{sn}: The Skew-Normal and Related Distributions Such as the
  Skew-t and the {SUN}}.
\newblock \proglang{R}~package version~2.0.2,
  \urlprefix\url{https://CRAN.R-project.org/package=sn}.

\bibitem[{Azzalini and Arellano-Valle(2013)}]{azzalini13}
Azzalini A, Arellano-Valle RB (2013).
\newblock \enquote{Maximum Penalized Likelihood Estimation for Skew-Normal and
  Skew-$t$ Distributions.}
\newblock \emph{Journal of Statistical Planning and Inference},
  \textbf{143}(2), 419--433.
\newblock \doi{10.1016/j.jspi.2012.06.022}.

\bibitem[{Bache and Wickham(2022)}]{magrittr}
Bache SM, Wickham H (2022).
\newblock \emph{\pkg{magrittr}: A Forward-Pipe Operator for \proglang{R}}.
\newblock \proglang{R}~package version~2.0.3,
  \urlprefix\url{https://CRAN.R-project.org/package=magrittr}.

\bibitem[{Baron and Kenny(1986)}]{baron86}
Baron RM, Kenny DA (1986).
\newblock \enquote{The Moderator-Mediator Variable Distinction in Social
  Psychological Research: Conceptual, Strategic, and Statistical
  Considerations.}
\newblock \emph{Journal of Personality and Social Psychology}, \textbf{51}(6),
  1173--1182.
\newblock \doi{10.1037/0022-3514.51.6.1173}.

\bibitem[{Blau(1977)}]{blau77}
Blau PM (1977).
\newblock \emph{Inequality and Heterogeneity: A Primitive Theory of Social
  Structure}.
\newblock Free Press, New York, NY.

\bibitem[{Bollen and Stine(1990)}]{bollen90}
Bollen KA, Stine R (1990).
\newblock \enquote{Direct and Indirect Effects: Classical and Bootstrap
  Estimates of Variability.}
\newblock \emph{Sociological Methodology}, \textbf{20}, 115--140.
\newblock \doi{10.2307/271084}.

\bibitem[{Canty and Ripley(2021)}]{boot}
Canty A, Ripley B (2021).
\newblock \emph{\pkg{boot}: Bootstrap \proglang{R} (\proglang{S-PLUS})
  Functions}.
\newblock \proglang{R}~package version~1.3-28,
  \urlprefix\url{https://CRAN.R-project.org/package=boot}.

\bibitem[{Carson \emph{et~al.}(2007)Carson, Tesluk, and Marrone}]{carson07}
Carson JB, Tesluk PE, Marrone JA (2007).
\newblock \enquote{Shared Leadership in Teams: An Investigation of Antecedent
  Conditions and Performance.}
\newblock \emph{Academy of Management Journal}, \textbf{50}(5), 1217--1234.
\newblock \doi{10.5465/amj.2007.20159921}.

\bibitem[{Cerin \emph{et~al.}(2009)Cerin, Leslie, Sugiyama, and Owen}]{cerin09}
Cerin E, Leslie E, Sugiyama T, Owen N (2009).
\newblock \enquote{Associations of Multiple Physical Activity Domains with
  Mental Well-Being.}
\newblock \emph{Mental Health and Physical Activity}, \textbf{2}(2), 55--64.
\newblock \doi{10.1016/j.mhpa.2009.09.004}.

\bibitem[{Chambers(1992)}]{chambers92}
Chambers JM (1992).
\newblock \enquote{Classes and Methods: Object-Oriented Programming in
  \proglang{S}.}
\newblock In JM~Chambers, TJ~Hastie (eds.), \emph{Statistical Models in
  \proglang{S}}, pp. 455--480. Chapman \& Hall/CRC, Boca Raton, FL.

\bibitem[{Chang \emph{et~al.}(2021)Chang, Cheng, Allaire, Sievert, Schloerke,
  Xie, Allen, McPherson, Dipert, and Borges}]{shiny}
Chang W, Cheng J, Allaire JJ, Sievert C, Schloerke B, Xie Y, Allen J, McPherson
  J, Dipert A, Borges B (2021).
\newblock \emph{\pkg{shiny}: Web Application Framework for \proglang{R}}.
\newblock \proglang{R}~package version~1.7.2,
  \urlprefix\url{https://CRAN.R-project.org/package=shiny}.

\bibitem[{Davison and Hinkley(1997)}]{davison97}
Davison AC, Hinkley DV (1997).
\newblock \emph{Bootstrap Methods and Their Application}.
\newblock Cambridge University Press, Cambridge, UK.

\bibitem[{Erreygers \emph{et~al.}(2018)Erreygers, Vandebosch, Vranjes,
  Baillien, and De~Witte}]{erreygers18}
Erreygers S, Vandebosch H, Vranjes I, Baillien E, De~Witte H (2018).
\newblock \enquote{The Interplay of Negative Experiences, Emotions and
  Affective Styles in Adolescents' Cybervictimization: A Moderated Mediation
  Analysis.}
\newblock \emph{Computers in Human Behavior}, \textbf{81}, 223--234.
\newblock \doi{10.1016/j.chb.2017.12.027}.

\bibitem[{Fox \emph{et~al.}(2022)Fox, Nie, and Byrnes}]{sem}
Fox J, Nie Z, Byrnes J (2022).
\newblock \emph{\pkg{sem}: Structural Equation Models}.
\newblock \proglang{R}~package version~3.1-15,
  \urlprefix\url{https://CRAN.R-project.org/package=sem}.

\bibitem[{Gaudiano \emph{et~al.}(2010)Gaudiano, Herbert, and
  Hayes}]{gaudiano10}
Gaudiano BA, Herbert JD, Hayes SC (2010).
\newblock \enquote{Is It the Symptom or the Relation to It? Investigating
  Potential Mediators of Change in Acceptance and Commitment Therapy for
  Psychosis.}
\newblock \emph{Behavior Therapy}, \textbf{41}(4), 543--554.
\newblock \doi{10.1016/j.beth.2010.03.001}.

\bibitem[{Hackman(1986)}]{hackman86}
Hackman JR (1986).
\newblock \enquote{The Psychology of Self-Management in Organizations.}
\newblock In MS~Pallack, RO~Perloff (eds.), \emph{Psychology and Work:
  Productivity, Change, and Employment}, pp. 89--136. American Psychological
  Association, Washington, DC.

\bibitem[{Harrison and Klein(2007)}]{harrison07}
Harrison DA, Klein KJ (2007).
\newblock \enquote{What's the Difference? Diversity Constructs as Separation,
  Variety, or Disparity in Organizations.}
\newblock \emph{Academy of Management Review}, \textbf{32}(4), 1199--1228.
\newblock \doi{10.5465/amr.2007.26586096}.

\bibitem[{Hayes(2018)}]{hayes18}
Hayes AF (2018).
\newblock \emph{Introduction to Mediation, Moderation, and Conditional Process
  Analysis}.
\newblock 2nd edition. The Guilford Press, New York, NY.

\bibitem[{{IBM Corp.}(2021)}]{SPSS}
{IBM Corp} (2021).
\newblock \emph{IBM \proglang{SPSS Statistics}, Version 28.0}.
\newblock Armonk, NY.
\newblock \urlprefix\url{https://www.ibm.com/products/spss-statistics}.

\bibitem[{Jehn(1995)}]{jehn95}
Jehn KA (1995).
\newblock \enquote{A Multimethod Examination of the Benefits and Detriments of
  Intragroup Conflict.}
\newblock \emph{Administrative Science Quarterly}, \textbf{40}(2), 256--285.
\newblock \doi{10.2307/2393638}.

\bibitem[{Judd and Kenny(1981)}]{judd81}
Judd CM, Kenny DA (1981).
\newblock \enquote{Process Analysis: Estimating Mediation in Treatment
  Evaluations.}
\newblock \emph{Evaluation Review}, \textbf{5}(5), 602--619.
\newblock \doi{10.1177/0193841x8100500502}.

\bibitem[{Kelley(2022)}]{MBESS}
Kelley K (2022).
\newblock \emph{\pkg{MBESS}: The \pkg{MBESS} \proglang{R} Package}.
\newblock \proglang{R}~package version~4.9.1,
  \urlprefix\url{https://CRAN.R-project.org/package=MBESS}.

\bibitem[{Koenker(2022)}]{quantreg}
Koenker R (2022).
\newblock \emph{\pkg{quantreg}: Quantile Regression}.
\newblock \proglang{R}~package version~5.93,
  \urlprefix\url{https://CRAN.R-project.org/package=quantreg}.

\bibitem[{Li and Cropanzano(2009)}]{li09}
Li A, Cropanzano R (2009).
\newblock \enquote{Fairness at the Group Level: Justice Climate and Intraunit
  Justice Climate.}
\newblock \emph{Journal of Management}, \textbf{35}(3), 564--599.
\newblock \doi{10.1177/0149206308330557}.

\bibitem[{Lindeman and Verkasalo(2005)}]{lindeman05}
Lindeman M, Verkasalo M (2005).
\newblock \enquote{Measuring Values With the Short Schwartz's Value Survey.}
\newblock \emph{Journal of Personality Assessment}, \textbf{85}(2), 170--178.
\newblock \doi{10.1207/s15327752jpa8502_09}.

\bibitem[{MacKinnon \emph{et~al.}(2002)MacKinnon, Lockwood, Hoffman, West, and
  Sheets}]{mackinnon02}
MacKinnon DP, Lockwood CM, Hoffman JM, West SG, Sheets V (2002).
\newblock \enquote{A Comparison of Methods to Test Mediation and Other
  Intervening Variable Effects.}
\newblock \emph{Psychological Methods}, \textbf{7}(1), 83--104.
\newblock \doi{10.1037/1082-989x.7.1.83}.

\bibitem[{MacKinnon \emph{et~al.}(2004)MacKinnon, Lockwood, and
  Williams}]{mackinnon04}
MacKinnon DP, Lockwood CM, Williams J (2004).
\newblock \enquote{Confidence Limits for the Indirect Effect: Distribution of
  the Product and Resampling Methods.}
\newblock \emph{Multivariate Behavioral Research}, \textbf{39}(1), 99--128.
\newblock \doi{10.1207/s15327906mbr3901_4}.

\bibitem[{MacKinnon \emph{et~al.}(1995)MacKinnon, Warsi, and
  Dwyer}]{mackinnon95}
MacKinnon DP, Warsi G, Dwyer JH (1995).
\newblock \enquote{A Simulation Study of Mediated Effect Measures.}
\newblock \emph{Multivariate Behavioral Research}, \textbf{30}(1), 41--62.
\newblock \doi{10.1207/s15327906mbr3001_3}.

\bibitem[{Maechler \emph{et~al.}(2022)Maechler, Rousseeuw, Croux, Todorov,
  Ruckstuhl, Salibi\'{a}n-Barrera, Verbeke, Koller, Concei\c{c}\~{a}o, and {Di
  Palma}}]{robustbase}
Maechler M, Rousseeuw P, Croux C, Todorov V, Ruckstuhl A, Salibi\'{a}n-Barrera
  M, Verbeke T, Koller M, Concei\c{c}\~{a}o ELT, {Di Palma} MA (2022).
\newblock \emph{\pkg{robustbase}: Basic Robust Statistics}.
\newblock \proglang{R}~package version~0.95.0,
  \urlprefix\url{https://CRAN.R-project.org/package=robustbase}.

\bibitem[{Mair and Wilcox(2020)}]{mair20}
Mair P, Wilcox R (2020).
\newblock \enquote{Robust Statistical Methods in \proglang{R} Using the
  \pkg{WRS2} Package.}
\newblock \emph{Behavior Research Methods}, \textbf{52}(2), 464--488.
\newblock \doi{10.3758/s13428-019-01246-w}.

\bibitem[{Mathieu and Rapp(2009)}]{mathieu09}
Mathieu JE, Rapp TL (2009).
\newblock \enquote{Laying the Foundation for Successful Team Performance
  Trajectories: The Roles of Team Charters and Performance Strategies.}
\newblock \emph{Journal of Applied Psychology}, \textbf{94}(1), 90--103.
\newblock \doi{10.1037/a0013257}.

\bibitem[{Mowday \emph{et~al.}(1979)Mowday, Steers, and Porter}]{mowday79}
Mowday RT, Steers RM, Porter LW (1979).
\newblock \enquote{The Measurement of Organizational Commitment.}
\newblock \emph{Journal of Vocational Behavior}, \textbf{14}(2), 224--247.
\newblock \doi{10.1016/0001-8791(79)90072-1}.

\bibitem[{Muller \emph{et~al.}(2005)Muller, Judd, and Yzerbyt}]{muller05}
Muller D, Judd CM, Yzerbyt VY (2005).
\newblock \enquote{When Moderation Is Mediated and Mediation Is Moderated.}
\newblock \emph{Journal of Personality and Social Psychology}, \textbf{89}(6),
  852--864.
\newblock \doi{10.1037/0022-3514.89.6.852}.

\bibitem[{Muth\'{e}n and Muth\'{e}n(2017)}]{Mplus}
Muth\'{e}n LK, Muth\'{e}n BO (2017).
\newblock \emph{\proglang{Mplus} User's Guide}.
\newblock Los Angeles, CA.
\newblock 8th edition, \urlprefix\url{https://www.statmodel.com/}.

\bibitem[{Pickett \emph{et~al.}(2012)Pickett, Yardley, and
  Kendrick}]{pickett12}
Pickett K, Yardley L, Kendrick T (2012).
\newblock \enquote{Physical Activity and Depression: A Multiple Mediation
  Analysis.}
\newblock \emph{Mental Health and Physical Activity}, \textbf{5}(2), 125--134.
\newblock \doi{10.1016/j.mhpa.2012.10.001}.

\bibitem[{Preacher and Hayes(2004)}]{preacher04}
Preacher KJ, Hayes AF (2004).
\newblock \enquote{\proglang{SPSS} and \proglang{SAS} Procedures for Estimating
  Indirect Effects in Simple Mediation Models.}
\newblock \emph{Behavior Research Methods, Instruments, \& Computers},
  \textbf{36}(4), 717--731.
\newblock \doi{10.3758/bf03206553}.

\bibitem[{Preacher and Hayes(2008)}]{preacher08}
Preacher KJ, Hayes AF (2008).
\newblock \enquote{Asymptotic and Resampling Strategies for Assessing and
  Computing Indirect Effects in Multiple Mediator Models.}
\newblock \emph{Behavior Research Methods}, \textbf{40}(3), 879--891.
\newblock \doi{10.3758/brm.40.3.879}.

\bibitem[{\proglang{SAS} {Institute Inc.}(2020)}]{SAS}
\proglang{SAS} {Institute Inc} (2020).
\newblock \emph{\proglang{SAS/STAT} Software, Version 15.2}.
\newblock Cary, NC.
\newblock \urlprefix\url{https://www.sas.com/}.

\bibitem[{{\proglang{R} Core Team}(2022)}]{R}
{\proglang{R} Core Team} (2022).
\newblock \emph{\proglang{R}: A Language and Environment for Statistical
  Computing}.
\newblock \proglang{R} Foundation for Statistical Computing, Vienna, Austria.
\newblock \urlprefix\url{https://www.R-project.org/}.

\bibitem[{Revelle(2022)}]{psych}
Revelle W (2022).
\newblock \emph{\pkg{psych}: Procedures for Psychological, Psychometric, and
  Personality Research}.
\newblock \proglang{R}~package version~2.2.5,
  \urlprefix\url{https://CRAN.R-project.org/package=psych}.

\bibitem[{Rosseel(2012)}]{rosseel12}
Rosseel Y (2012).
\newblock \enquote{\pkg{lavaan}: An \proglang{R} Package for Structural
  Equation Modeling.}
\newblock \emph{Journal of Statistical Software}, \textbf{48}(2), 1--36.
\newblock \doi{10.18637/jss.v048.i02}.

\bibitem[{Rousseeuw and Yohai(1984)}]{rousseeuw84}
Rousseeuw PJ, Yohai VJ (1984).
\newblock \enquote{Robust Regression by Means of {S}-Estimators.}
\newblock In J~Franke, W~H\"{a}rdle, D~Martin (eds.), \emph{Robust and
  Nonlinear Time Series Analysis}, volume~26 of \emph{Lecture Notes in
  Statistics}, pp. 256--272. Springer-Verlag, New York, NY.

\bibitem[{Salibi\'{a}n-Barrera and Van~Aelst(2008)}]{salibian08}
Salibi\'{a}n-Barrera M, Van~Aelst S (2008).
\newblock \enquote{Robust Model Selection Using Fast and Robust Bootstrap.}
\newblock \emph{Computational Statistics \& Data Analysis}, \textbf{52}(12),
  5121--5135.
\newblock \doi{10.1016/j.csda.2008.05.007}.

\bibitem[{Salibi\'{a}n-Barrera and Yohai(2006)}]{salibian06}
Salibi\'{a}n-Barrera M, Yohai VJ (2006).
\newblock \enquote{A Fast Algorithm for {S}-Regression Estimates.}
\newblock \emph{Journal of Computational and Graphical Statistics},
  \textbf{15}(2), 414--427.
\newblock \doi{10.1198/106186006x113629}.

\bibitem[{Salibi\'{a}n-Barrera and Zamar(2002)}]{salibian02}
Salibi\'{a}n-Barrera M, Zamar RH (2002).
\newblock \enquote{Bootstrapping Robust Estimates of Regression.}
\newblock \emph{The Annals of Statistics}, \textbf{30}(2), 556--582.
\newblock \doi{10.1214/aos/1021379865}.

\bibitem[{Schwarz(1978)}]{schwarz78}
Schwarz G (1978).
\newblock \enquote{Estimating the Dimension of a Model.}
\newblock \emph{The Annals of Statistics}, \textbf{6}(2), 461--464.
\newblock \doi{10.1214/aos/1176344136}.

\bibitem[{Shrout and Bolger(2002)}]{shrout02}
Shrout PE, Bolger N (2002).
\newblock \enquote{Mediation in Experimental and Nonexperimental Studies: New
  Procedures and Recommendations.}
\newblock \emph{Psychological Methods}, \textbf{7}(4), 422--445.
\newblock \doi{10.1037/1082-989x.7.4.422}.

\bibitem[{Sobel(1982)}]{sobel82}
Sobel ME (1982).
\newblock \enquote{Asymptotic Confidence Intervals for Indirect Effects in
  Structural Equation Models.}
\newblock \emph{Sociological Methodology}, \textbf{13}, 290--312.
\newblock \doi{10.2307/270723}.

\bibitem[{Steen \emph{et~al.}(2017)Steen, Loeys, Moerkerke, and
  Vansteelandt}]{steen17}
Steen J, Loeys T, Moerkerke B, Vansteelandt S (2017).
\newblock \enquote{\pkg{medflex}: An \proglang{R} Package for Flexible
  Mediation Analysis Using Natural Effect Models.}
\newblock \emph{Journal of Statistical Software}, \textbf{76}(11), 1--46.
\newblock \doi{10.18637/jss.v076.i11}.

\bibitem[{Tingley \emph{et~al.}(2014)Tingley, Yamamoto, Hirose, Keele, and
  Imai}]{tingley14}
Tingley D, Yamamoto T, Hirose K, Keele L, Imai K (2014).
\newblock \enquote{\pkg{mediation}: \proglang{R} Package for Causal Mediation
  Analysis.}
\newblock \emph{Journal of Statistical Software}, \textbf{59}(5), 1--38.
\newblock \doi{10.18637/jss.v059.i05}.

\bibitem[{Vuorre(2021)}]{bmlm}
Vuorre M (2021).
\newblock \emph{\pkg{bmlm}: Bayesian Multilevel Mediation}.
\newblock \proglang{R}~package version~1.3.12,
  \urlprefix\url{https://CRAN.R-project.org/package=bmlm}.

\bibitem[{Wickham(2011)}]{wickham11}
Wickham H (2011).
\newblock \enquote{\pkg{testthat}: Get Started with Testing.}
\newblock \emph{The \proglang{R} Journal}, \textbf{3}(1), 5--10.
\newblock \doi{10.32614/rj-2011-002}.

\bibitem[{Wickham(2016)}]{wickham16}
Wickham H (2016).
\newblock \emph{\pkg{ggplot2}: Elegant Graphics for Data Analysis}.
\newblock Springer-Verlag, New York, NY.
\newblock \urlprefix\url{https://ggplot2.tidyverse.org}.

\bibitem[{Wood \emph{et~al.}(2008)Wood, Goodman, Beckmann, and Cook}]{wood08}
Wood RE, Goodman JS, Beckmann N, Cook A (2008).
\newblock \enquote{Mediation Testing in Management Research: A Review and
  Proposals.}
\newblock \emph{Organizational Research Methods}, \textbf{11}(2), 270--295.
\newblock \doi{10.1177/1094428106297811}.

\bibitem[{Yohai(1987)}]{yohai87}
Yohai VJ (1987).
\newblock \enquote{High Breakdown-Point and High Efficiency Robust Estimates
  for Regression.}
\newblock \emph{The Annals of Statistics}, \textbf{15}(2), 642--656.
\newblock \doi{10.1214/aos/1176350366}.

\bibitem[{Yu and Li(2022)}]{mma}
Yu Q, Li B (2022).
\newblock \emph{\pkg{mma}: Multiple Mediation Analysis}.
\newblock \proglang{R}~package version~10.6-1,
  \urlprefix\url{https://CRAN.R-project.org/package=mma}.

\bibitem[{Yuan and MacKinnon(2014)}]{yuan14}
Yuan Y, MacKinnon DP (2014).
\newblock \enquote{Robust Mediation Analysis Based on Median Regression.}
\newblock \emph{Psychological Methods}, \textbf{19}(1), 1--20.
\newblock \doi{10.1037/a0033820}.

\bibitem[{Zhao \emph{et~al.}(2010)Zhao, Lynch, and Chen}]{zhao10}
Zhao X, Lynch JGJ, Chen Q (2010).
\newblock \enquote{Reconsidering Baron and Kenny: Myths and Truths about
  Mediation Analysis.}
\newblock \emph{Journal of Consumer Research}, \textbf{37}(2), 197--206.
\newblock \doi{10.1086/651257}.

\bibitem[{Zu and Yuan(2010)}]{zu10}
Zu J, Yuan KH (2010).
\newblock \enquote{Local Influence and Robust Procedures for Mediation
  Analysis.}
\newblock \emph{Multivariate Behavioral Research}, \textbf{45}(1), 1--44.
\newblock \doi{10.1080/00273170903504695}.

\end{thebibliography}


\newpage

\begin{appendix}


\section{Details on the diagnostic plot with a tolerance ellipse}
\label{app:ellipse}

The diagnostic plot from function \code{ellipse_plot()} exploits the
relationship between regression coefficients and the covariance matrix to
draw a tolerance ellipse that illustrates how well the regression results
represent the data.  Our recommended robust bootstrap test for mediation
analysis is based on a robust regression estimator that assigns robustness
weights to the observations.  Those robustness weights lie between 0 and 1,
with lower weights indicating a higher degree of deviation.  The corresponding
tolerance ellipse is computed based on a weighted sample covariance matrix,
using the weights returned by the robust regression.  However, for such a plot
to be meaningful, the weighted sample covariance matrix needs to yield a
Fisher consistent estimator of the true covariance matrix under the model
distribution.

The diagnostic plot is most useful if the data come from a multivariate
elliptical distribution.  Consider the regression model
\begin{equation*}
Y = \alpha + \vect{X}^\top \vect{\beta} + \sigma \varepsilon,
\end{equation*}
In addition to the usual assumptions that $\varepsilon \sim N(0, 1)$ and that
$\vect{X}$ and $\varepsilon$ are uncorrelated, we therefore also assume
for the diagnostic plot that $\vect{X} \sim N(\vect{\mu}_{\vect{X}},
\mat{\Sigma}_{\vect{X}\vect{X}})$.  We emphasize that this additional
assumption is made only for the diagnostic plot, it is not required for
the general applicability of our robust bootstrap test.  Then we have
$Z = (Y, \vect{X}^\top)^\top \sim N(\vect{\mu}, \mat{\Sigma})$ with
\begin{equation*}
\vect{\mu} =
\left(
\begin{array}{c}
\mu_{Y} \\
\vect{\mu}_{\vect{X}}
\end{array}
\right)
\qquad \text{and} \qquad
\mat{\Sigma} =
\left(
\begin{array}{cc}
\Sigma_{YY}               & \vect{\Sigma}_{Y\vect{X}} \\
\vect{\Sigma}_{\vect{X}Y} & \mat{\Sigma}_{\vect{X}\vect{X}}
\end{array}
\right).
\end{equation*}
Furthermore, the regression coefficients can be written as
\begin{align*}
\beta  &= \mat{\Sigma}_{\vect{X}\vect{X}}^{-1} \mat{\Sigma}_{\vect{X}Y}, \\
\alpha &= \mu_{Y} - \vect{\mu}_{\vect{X}}^\top \vect{\beta}.
\end{align*}

With observations $\obs{z}_{i} = (y_{i}, \obs{x}_{i}^\top)^\top$ and with with
robustness weights $w_{i}$ from robust regression, $i = 1, \dots, n$, we define
the weighted center estimate $\obs{m}$ and the weighted scatter matrix
$\mat{S}$ as
\begin{align*}
\obs{m} &=
\frac{1}{\sum_{i = 1}^{n} w_{i}}
\sum_{i = 1}^{n} w_{i} \obs{z}_{i}, \\
\mat{S} &=
\frac{1}{\left( \sum_{i = 1}^{n} w_{i} \right) - 1}
\sum_{i = 1}^{n} w_{i} (\obs{z}_{i} - \obs{m}) (\obs{z}_{i} - \obs{m})^\top.
\end{align*}
Note that the denominator of $\mat{S}$ is chosen such that the weighted scatter
matrix reduces to the unbiased sample covariance matrix if all observations
receive full weight $w_{i} = 1$, \mbox{$i = 1, \dots, n$}. Let $F$ denote the
cumulative distribution function (CDF) of the multivariate normal distribution
$N(\vect{\mu}, \mat{\Sigma})$ and let $\Phi$ denote the CDF of the univariate
standard normal distribution $N(0, 1)$.  Then the functional forms of the
estimators are given by
\begin{align*}
\vect{m}(F) &=
\frac{1}{\delta} \int w(\varepsilon) \vect{z} \: dF(\vect{z}), \\
\mat{S}(F) &=
\frac{1}{\delta}
\int w(\varepsilon) (\vect{z} - \vect{m}(F)) (\vect{z} - \vect{m}(F))^\top \:
dF(\vect{z}),
\end{align*}
where $\delta = \int w(\varepsilon) \: d\Phi(\varepsilon)$.

Keep in mind that we only consider robust regression with symmetric loss
functions such that the weight function $w$ is also symmetric, and let
$\vect{m}(F) = (m_{Y}(F), \vect{m}_{\vect{X}}(F)^\top)^\top$.  For the
explanatory variables $\vect{X}$, we have
\begin{equation*}
\vect{m}_{\vect{X}}(F) =
\frac{1}{\delta} \int w(\varepsilon) \vect{x} \: dF(\vect{z}) =
\frac{1}{\delta}
\underbrace{\int w(\varepsilon) \: d\Phi(\varepsilon)}_{=\delta}
\int \vect{x} \: dF_{\vect{X}}(\vect{x}) = \vect{\mu}_{\vect{X}},
\end{equation*}
where $F_{\vect{X}}$ denotes the joint CDF of $\vect{X}$.  To show that
$m_{Y}(F) = \mu_{Y}$, we can assume without loss of generality that
$\mu = (0, \dots, 0)^\top$ and that $\vect{\beta} = (0, \dots, 0)^\top$.  Then
$\alpha = 0$ and $\varepsilon = Y / \sigma$, and we need to show that
$m_{Y}(F) = 0$.  We obtain
\begin{equation} \label{eq:mean_Y}
m_{Y}(F) =
\frac{1}{\delta} \int w(\varepsilon) y \: dF(\vect{z}) =
\frac{1}{\delta} \int_{-\infty}^{\infty}
\underbrace{w \left( \frac{y}{\sigma} \right) y f_{Y}(y)}_{=g(y)}
\: dy,
\end{equation}
where $f_{Y}$ denotes the probability density function of the marginal
distribution of $Y$.  For the integrand in \eqref{eq:mean_Y}, it holds that
$g(-y) = -g(y)$, hence we have $m_{Y}(F) = 0$.  Thus $\vect{\hat{\mu}} =
\vect{m}$ is Fisher consistent for $\vect{\mu}$.

Since $0 \leq w(\varepsilon) \leq 1$, $\mat{S}(F)$ is expected to underestimate
(some elements of) the covariance matrix $\mat{\Sigma}$.  However, for the
submatrix involving only the explanatory variables $\vect{X}$, we obtain
\begin{align*}
\mat{S}_{\vect{X}\vect{X}}(F)
&= \frac{1}{\delta}
   \int w(\varepsilon)
   (\vect{x} - \underbrace{\vect{m}_{\vect{X}}(F)}_{=\vect{\mu}_{\vect{X}}})
   (\vect{x} - \underbrace{\vect{m}_{\vect{X}}(F)}_{=\vect{\mu}_{\vect{X}}})^\top
   \: dF(\vect{z}) \\
&= \frac{1}{\delta}
   \underbrace{\int w(\varepsilon) \: d\Phi(\varepsilon)}_{=\delta}
   \int (\vect{x} - \vect{\mu}_{\vect{X}})
   (\vect{x} - \vect{\mu}_{\vect{X}})^\top \: dF_{\vect{X}}(\vect{x}) =
   \mat{\Sigma}_{\vect{X}\vect{X}}.
\end{align*}
Thus
\begin{equation*}
\mat{\hat{\Sigma}}_{\vect{X}\vect{X}} = \mat{S}_{\vect{X}\vect{X}} =
\frac{1}{\left( \sum_{i = 1}^{n} w_{i} \right) - 1}
\sum_{i = 1}^{n} w_{i} (\obs{x}_{i} - \obs{\bar{x}})
(\obs{x}_{i} - \obs{\bar{x}})^\top
\end{equation*}
is Fisher consistent for $\mat{\Sigma}_{\vect{X}\vect{X}}$.  With Fisher
consistent estimators $\vect{\hat{\beta}}$ and $\hat{\sigma}$ from robust
regression, we can therefore construct a Fisher consistent estimator
$\mat{\hat{\Sigma}}$ of the covariance matrix $\mat{\Sigma}$ by computing
\begin{align*}
\vect{\hat{\Sigma}}_{\vect{X}Y} &=
\mat{\hat{\Sigma}}_{\vect{X}\vect{X}} \vect{\hat{\beta}}, \\
\hat{\Sigma}_{YY} &=
\vect{\hat{\beta}}^\top \mat{\hat{\Sigma}}_{\vect{X}\vect{X}}
\vect{\hat{\beta}} + \hat{\sigma}^{2}, \\
\mat{\hat{\Sigma}} &=
\left(
\begin{array}{cc}
\hat{\Sigma}_{YY}               & \vect{\hat{\Sigma}}_{\vect{X}Y}^\top \\
\vect{\hat{\Sigma}}_{\vect{X}Y} & \mat{\hat{\Sigma}}_{\vect{X}\vect{X}}
\end{array}
\right)
.
\end{align*}

\end{appendix}


\section{Additional summary output of robust bootstrap tests}
\label{app:output}

We first produce the summary of the robust bootstrap test for the serial
multiple mediator model from Section~\ref{sec:serial}.  Since the diagnostic
plot for this example is already shown in Figure~\ref{fig:weight}, we set the
argument \code{plot = FALSE} to suppress the diagnostic plot.
\begin{Schunk}
\begin{Sinput}
R> summary(robust_boot_serial, plot = FALSE)
\end{Sinput}
\begin{Soutput}
Robust bootstrap tests for indirect effects via regression

Serial multiple mediator model

x  = ValueDiversity
y  = TeamScore
m1 = TaskConflict
m2 = TeamCommitment

Sample size: 89
---
Outcome variable: TaskConflict

Coefficients:
                 Data   Boot Std. Error z value Pr(>|z|)    
(Intercept)    1.1182 1.1171     0.1794   6.226 4.77e-10 ***
ValueDiversity 0.3197 0.3208     0.1071   2.994  0.00275 ** 

Robust residual standard error: 0.3033 on 87 degrees of freedom
Robust R-squared:  0.1181,	Adjusted robust R-squared:  0.108
Robust F-statistic: 9.113 on 1 and Inf DF,  p-value: 0.002539

Robustness weights:
4 observations are potential outliers with weight <= 1.3e-05:
[1] 48 58 76 79
---
Outcome variable: TeamCommitment

Coefficients:
                   Data     Boot Std. Error z value Pr(>|z|)    
(Intercept)     4.33385  4.33709    0.34436  12.595   <2e-16 ***
TaskConflict   -0.33659 -0.33722    0.17763  -1.898   0.0576 .  
ValueDiversity  0.06523  0.06319    0.18722   0.338   0.7357    

Robust residual standard error: 0.3899 on 86 degrees of freedom
Robust R-squared:  0.08994,	Adjusted robust R-squared:  0.06878
Robust F-statistic: 1.497 on 2 and Inf DF,  p-value: 0.2239

Robustness weights:
Observation 6 is a potential outlier with weight 0
---
Outcome variable: TeamScore

Coefficients:
                  Data    Boot Std. Error z value Pr(>|z|)    
(Intercept)    61.9660 60.7349    16.4148   3.700 0.000216 ***
TaskConflict    0.3240  0.1877     2.7909   0.067 0.946369    
TeamCommitment  9.2138  9.4999     4.5906   2.069 0.038508 *  
ValueDiversity  0.2024  0.4193     3.0612   0.137 0.891065    

Robust residual standard error: 8.42 on 85 degrees of freedom
Robust R-squared:  0.1746,	Adjusted robust R-squared:  0.1455
Robust F-statistic: 1.194 on 3 and Inf DF,  p-value: 0.3103

Robustness weights:
3 observations are potential outliers with weight <= 0:
[1] 10 32 38
---
Total effect of x on y:
                   Data     Boot Std. Error z value Pr(>|z|)
ValueDiversity -0.08467 -0.04260    3.28764  -0.013     0.99

Direct effect of x on y:
                 Data   Boot Std. Error z value Pr(>|z|)
ValueDiversity 0.2024 0.4193     3.0612   0.137    0.891

Indirect effects of x on y:
             Data     Boot  Lower    Upper
Total     -0.2870 -0.46185 -6.859  2.87613
Indirect1  0.1036  0.07264 -1.668  2.24594
Indirect2  0.6010  0.45184 -3.130  5.13150
Indirect3 -0.9916 -0.98633 -3.909 -0.06841

Indirect effect paths:
 Label     Path                                                           
 Indirect1 ValueDiversity -> TaskConflict   -> TeamScore                  
 Indirect2 ValueDiversity -> TeamCommitment -> TeamScore                  
 Indirect3 ValueDiversity -> TaskConflict   -> TeamCommitment -> TeamScore
---
Level of confidence: 95 %

Number of bootstrap replicates: 5000
---
Signif. codes:  0 '***' 0.001 '**' 0.01 '*' 0.05 '.' 0.1 ' ' 1
\end{Soutput}
\end{Schunk}

\bigskip
Finally, we display the summary of the robust bootstrap test for the parallel
multiple mediator model with control variables from Section~\ref{sec:parallel}.
Figure~\ref{fig:summary-parallel} contains the resulting diagnostic plot.
\begin{Schunk}
\begin{Sinput}
R> summary(robust_boot_parallel)
\end{Sinput}
\begin{Soutput}
Robust bootstrap tests for indirect effects via regression

Parallel multiple mediator model

x  = SharedLeadership
y  = TeamPerformance
m1 = ProceduralJustice
m2 = InteractionalJustice

Covariates:
[1] AgeDiversity    GenderDiversity

Sample size: 89
---
Outcome variable: ProceduralJustice

Coefficients:
                    Data    Boot Std. Error z value Pr(>|z|)    
(Intercept)      3.37132 3.34898    0.22221  15.071   <2e-16 ***
SharedLeadership 0.06287 0.06732    0.03632   1.853   0.0638 .  
AgeDiversity     0.05232 0.04914    0.03020   1.627   0.1037    
GenderDiversity  0.22030 0.20656    0.14977   1.379   0.1678    

Robust residual standard error: 0.2377 on 85 degrees of freedom
Robust R-squared:  0.1277,	Adjusted robust R-squared:  0.09696
Robust F-statistic: 1.208 on 3 and Inf DF,  p-value: 0.3052

Robustness weights:
No potential outliers with weight < 0.0011 detected.
The minimum weight is 0.27.
---
Outcome variable: InteractionalJustice

Coefficients:
                      Data      Boot Std. Error z value Pr(>|z|)    
(Intercept)       3.276732  3.261434   0.261999  12.448  < 2e-16 ***
SharedLeadership  0.151376  0.153656   0.034581   4.443 8.85e-06 ***
AgeDiversity     -0.007949 -0.005299   0.050795  -0.104    0.917    
GenderDiversity   0.344611  0.338612   0.184200   1.838    0.066 .  

Robust residual standard error: 0.2788 on 85 degrees of freedom
Robust R-squared:  0.2847,	Adjusted robust R-squared:  0.2595
Robust F-statistic: 2.169 on 3 and Inf DF,  p-value: 0.08934

Robustness weights:
2 observations are potential outliers with weight <= 0:
[1] 31 57
---
Outcome variable: TeamPerformance

Coefficients:
                           Data       Boot Std. Error z value Pr(>|z|)   
(Intercept)          -0.8007568 -0.8296269  0.6028351  -1.376  0.16876   
ProceduralJustice     0.6335972  0.6527230  0.2007240   3.252  0.00115 **
InteractionalJustice  0.5354262  0.5247245  0.1754996   2.990  0.00279 **
SharedLeadership     -0.0164857 -0.0157296  0.0504240  -0.312  0.75508   
AgeDiversity          0.0025378  0.0008948  0.0362812   0.025  0.98032   
GenderDiversity       0.2756176  0.2716191  0.1549740   1.753  0.07966 . 

Robust residual standard error: 0.2558 on 83 degrees of freedom
Robust R-squared:  0.5988,	Adjusted robust R-squared:  0.5746
Robust F-statistic: 3.476 on 5 and Inf DF,  p-value: 0.003833

Robustness weights:
3 observations are potential outliers with weight <= 0:
[1] 33 38 79
---
Total effect of x on y:
                    Data    Boot Std. Error z value Pr(>|z|)  
SharedLeadership 0.10440 0.10973    0.05295   2.072   0.0382 *

Direct effect of x on y:
                     Data     Boot Std. Error z value Pr(>|z|)
SharedLeadership -0.01649 -0.01573    0.05042  -0.312    0.755

Indirect effects of x on y:
                        Data    Boot     Lower  Upper
Total                0.12089 0.12546 4.398e-02 0.2164
ProceduralJustice    0.03984 0.04568 6.206e-05 0.1184
InteractionalJustice 0.08105 0.07978 3.022e-02 0.1538
---
Level of confidence: 95 %

Number of bootstrap replicates: 5000
---
Signif. codes:  0 '***' 0.001 '**' 0.01 '*' 0.05 '.' 0.1 ' ' 1
\end{Soutput}
\begin{figure}[t!]

{\centering \includegraphics[width=0.7\textwidth]{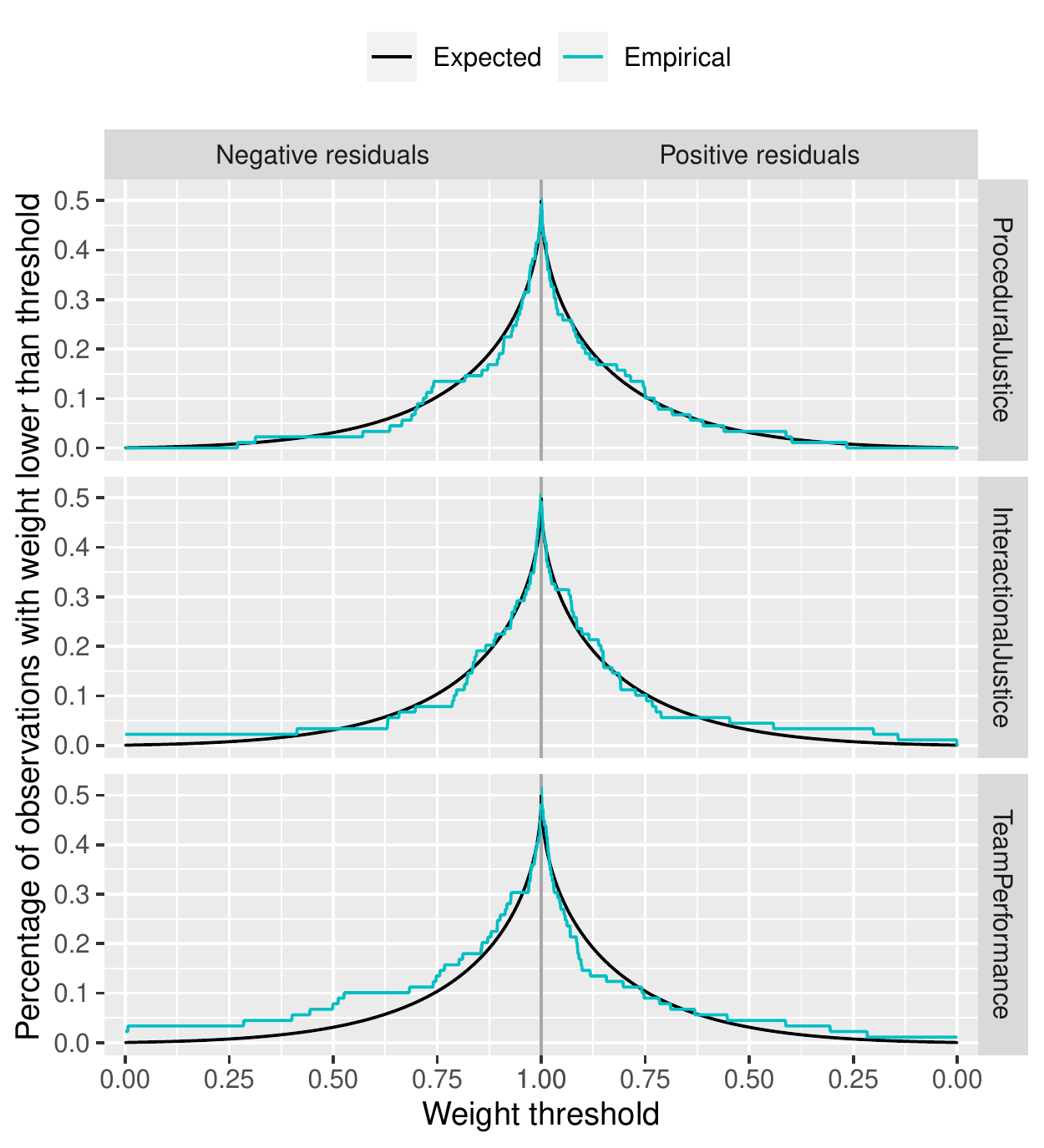} 

}

\caption[Diagnostic plot of the regression weights from the robust bootstrap procedure of \citet{alfons22a} in the example for a parallel multiple mediator model]{Diagnostic plot of the regression weights from the robust bootstrap procedure of \citet{alfons22a} in the example for a parallel multiple mediator model.}\label{fig:summary-parallel}
\end{figure}
\end{Schunk}
\newpage  


\end{document}